# Novel Machine Learning Approaches for Improving the Reproducibility and Reliability of Functional and Effective Connectivity from Functional MRI

**Cooper J. Mellema, PhD**[1,2,5]
**Albert A. Montillo, PhD**[1,2,3,4,5]
[1]Lyda Hill Department of Bioinformatics
[2]Biomedical Engineering Department
[3]Advanced Imaging Research Center
[4]Radiology Department
[5]University of Texas Southwestern Medical Center
Cooper.Mellema@UTSouthwestern.edu
Albert.Montillo@UTSouthwestern.edu

**Abstract**

***Objective:*** New measures of human brain connectivity are needed to address gaps in the existing measures and facilitate the study of brain function, cognitive capacity, and identify early markers of human disease. Traditional approaches to measure functional connectivity (FC) between pairs of brain regions in functional MRI (fMRI), such as correlation and partial correlation, fail to capture nonlinear aspects in the regional associations. We propose a new machine learning based measure of functional connectivity (*ML.FC*) which efficiently captures linear and nonlinear aspects.

***Approach:*** To capture directed information flow between brain regions, effective connectivity (EC) metrics, including dynamic causal modeling (DCM) and structural equation modeling (SEM) have been used. However, these methods are impractical to compute across the many regions of the whole brain. Therefore, we propose two new EC measures. The first, a machine learning based measure of effective connectivity (*ML.EC*), measures nonlinear aspects across the entire brain. The second, Structurally Projected Granger Causality (*SP.GC*) adapts Granger Causal connectivity to efficiently characterize and regularize the whole brain EC connectome to respect underlying biological structural connectivity. The proposed measures are compared to traditional measures in terms of *reproducibility* and the *ability to predict individual traits* in order to demonstrate these measures' internal validity. We use four repeat scans of the same individuals from the Human Connectome Project (HCP) and measure the ability of the measures to predict individual subject physiologic and cognitive traits.

***Main results:*** The proposed new FC measure of *ML.FC* attains high reproducibility (mean intra-subject $R^2$ of 0.44), while the proposed EC measure of *SP.GC* attains the highest predictive power (mean $R^2$ across prediction tasks of 0.66).

***Significance:*** The proposed methods are highly suitable for achieving high reproducibility and predictiveness and demonstrate their strong potential for future neuroimaging studies.

## 1 Introduction

The connectivity of the human brain is integral to cognitive capacity, can be an early marker for human disease, and underlies the fundamental functioning of the central nervous system (Ashburner et al., 2004). However, measuring connectivity *in vivo* has proven problematic (Andellini et al., 2015; Fiecas et al., 2013; Rowe, 2010). Functional magnetic resonance imaging (fMRI[1]) of the brain measures the blood-oxygen-level-dependent (BOLD) signal and serves as an indirect measure of neural activity. The brain scan can be parcellated into neuroanatomical regions and the mean regional time series can be computed from the voxels in each region. By measuring temporal relationships between the mean BOLD signal from two or more regions of the brain, the underlying direct and indirect connectivity and communication within the brain can be probed. The connections between regions can then be used to represent the subject-specific connectome as a connectivity graph with each region represented as a node in the graph, while the edges between nodes are assigned an edge strength proportion to the pairwise regional connectivity.

[1] *All abbreviations used in this manuscript are described in detail in* ***Supplemental Table S1***

Connectivity measures are calculated from fMRI using a measure of similarity or information transfer between the mean regional BOLD timeseries of a pair of regions. Connectivity metrics can be grouped into undirected functional connectivity (FC) metrics and directed effective connectivity (EC) metrics. *Functional connectivity* is defined as the temporal coincidence of spatially distant neurophysiological events (Ashburner et al., 2004) and it has been used to characterize the human connectome in both health and disease (Cohen et al., 2017; Smitha et al., 2017). FC is traditionally calculated as the correlation or partial correlation between the regional timeseries. Meanwhile, *effective connectivity* is defined as the influence one neural system exerts over another (Ashburner et al., 2004). Broadly, this is a model-dependent measure wherein the information transfer between mean regional timeseries is quantified from the goodness of fit of a model that predicts one of the timeseries from one or more of the other timeseries. Examples of EC measures include Granger causality (Abidin et al., 2019; Chockanathan et al., 2019; Granger, 1969; Spencer et al., 2018), dynamic causal modeling (Friston et al., 2019; Park et al., 2018), and structured equation modeling (Rowe, 2010), which have been widely deployed for connectome characterization. EC is inherently directional as it captures the direction of information flow over time (Bielczyk et al., 2019). EC is model-dependent and requires more computation than FC but suppresses spurious indirect connections and identifies linkages that are potentially causal and not simply correlated.

Traditional FC and EC measures have several limitations, for which we propose new solutions. Common FC measures include Pearson's *r*, partial correlation, and *spectral* Granger Causality (Mingzhou Ding, Yongheng Chen, Stephen L. Bressler, 2006). Each of these methods measure a degree of linear association between two mean regional timeseries; however, the actual relationship between mean regional brain activity is nonlinear (Friston et al., 2019). Therefore, we propose nonlinear machine learning models that measure functional connectivity while capturing such nonlinearities. Of the different EC measures, we focus on Granger causal (GC) methods as they are data-driven approaches that can be used when many neuroanatomical regions, *N*, are to be analyzed (e.g., $N$>50). In modern fMRI connectivity analysis *N* is often a hundred or more. Alternative causal models, including dynamic causal models (DCM) and structured equation models (SEM), typically apply an exhaustive search over possible connectivity patterns, making analysis at this ROI granularity intractable for current compute hardware. Limiting the connectivity to a subset of the brain, such as intra-DMN connectivity, is often used as a workaround, but this restricts the portion of the brain under consideration and can miss important interactions (Friston et al., 2019; Rowe, 2010). Granger causal methods have limitations as well but these, we hypothesize, are surmountable, including: model selection procedures, regularization, scalability (the traditional GC method requires fitting $O(N^2)$ sub-models where *N* is the number of regions under analysis), an inability to capture nonlinear interactions, and the absence of the incorporation of prior knowledge of brain architecture (Ashburner et al., 2004). To address these limitations, we propose two measures of effective connectivity. The first measure which we call Machine-Learning Functional Connectivity ($\boldsymbol{ML.FC}$), uses a nonlinear machine learning model to quantify nonlinear pairwise timeseries associations. Our method is more scalable because the number of required models to fit scales as $O(N)$. Our second measure, which we call Structurally-Projected Granger Causality ($\boldsymbol{SP.GC}$), reformulates Granger causal connectivity in two ways. First, we regularize the connectivity computation using a structural connectivity prior derived from diffusion MRI. Streamline tractography is performed on diffusion MRI from the human connectome project (HCP) and a streamline atlas is generated (Yeh et al., 2018). The log of the number of streamlines connecting regions is used as a measure of pairwise *structural* connectivity. This is used to regularize the functional interactions inferred between regional timeseries via a tradeoff between the raw functional data interactivity and fiber bundle connectivity. As actual neural communication occurs through physical connections, this constraint is a natural choice of a prior to guide brain functional connectivity (Allen and Weylandt, 2019; Dillon et al., 2017; Huang and Ding, 2016; Maglanoc et al., 2020; Manning et al., 2018). The second way we reformulate Granger causal connectivity is to perform dimensionality reduction. Calculating the connectivity in a low dimensional space affords several advantages including: simplifying model optimization as there are fewer weights to tune and providing further regularization to stabilize fMRI interpretation and increase reproducibility. This dimensionality reduction is achieved by projecting the mean regional timeseries into a low dimensional space informed by the streamline structural connectivity prior. Each of our proposed measures is evaluated for reproducibility and the ability to predict cognitive and physiological traits of the HCP participants in our study.

A connectivity measure should produce a similar connectivity matrix for a given individual across repeat fMRI scans that are acquired within a short window of time. Therefore, we evaluated the proposed FC and EC measures reproducibility across four repeated fMRI scans of each individual in our Human Connectome Project (HCP)-derived dataset. A reproducible measure better characterizes an individual's connectivity fingerprint and is therefore more useful to capture true differences between individuals (Noble et al., 2019; Waller et al., 2017). Reproducibility is necessary, but insufficient to show that the proposed measures have validity; therefore we also measure the predictive power of each FC and EC metric in three relevant domains: a purely physiological domain predicting mean arterial pressure, a purely cognitive domain measuring fluid intelligence, and a combined physiologic and cognitive domain measuring stress. These were chosen as representative targets of interest of researchers and clinicians interested in predictions for physiology (e.g., stroke, aging), cognition (e.g., memory, PTSD), or a combination of the two (e.g., stress, neurodegeneration) for diagnoses and treatment. Measures that are both reproducible and have consistently high predictive power across multiple tasks are significantly more useful as candidate biomarkers (Noble et al., 2017b; Noble et al., 2017a; Termenon et al., 2016; Waller et al., 2017). We postulate that a measure that is both more reproducible and predictive is a better representation of true underlying neural patterns than alternative measures. The contributions of this work are: (1) the development of a new functional connectivity metric (*ML.FC*) and a new effective connectivity metric (*ML.EC*) that efficiently capture nonlinear associations between brain regions, (2) the development of a new effective connectivity metric (*SP.GC*) that incorporates a structural connectivity prior while efficiently measuring associations across all brain regions in a low dimensional space, (3) a quantitative comparison of the proposed measures to traditional measures of connectivity in terms of reproducibility and the power to predictive traits of individual subjects. Finally, (4) we recommend individual measures that hold the most potential to advance the study of human brain connectivity in health and disease based on the quantitative comparison.

# 2 Methods and Materials

## 2.1 Methods

### 2.1.1 Proposed machine learning-based functional connectivity ($ML.FC$) measures

Characterizing brain connectivity to better understand both health and disease is a complex process requiring measuring both linear and nonlinear aspects of information transfer between brain regions.

Classical means of performing this characterization include the use of Pearson's *r*, partial correlation, and spectral Granger causality. (For definitions of classical measures of FC, see **Supplemental Section 9.1.2**.) Central to this premise, we propose the construction of a machine learning model to calculate functional connectivity, an approach we denote as $\boldsymbol{ML.FC}$. This model predicts the activity, $a$, at a given node $j$ by using the information present at all other nodes (brain regions) at any given time, $t$. As illustrated in **Equation 1**, we use a nonlinear model $M$ to predict the activity at region $j$ at time $t$ from all other regions under analysis $R$ except region $j$.

$$a_{j,t} = M\left(\vec{a}_{i\in R\setminus j,t}\right) + \varepsilon_{j,t}, \qquad a = activity, \qquad \varepsilon = error \qquad \textbf{(1)}$$

This model simultaneously learns the association between all other nodes' activity and the target node $j$. The weight assigned to each covariate *quantifies* the amount of information the model is using from that node to predict the target node $j$, which is a putative measure of the *connectivity* between each node $i$ and $j$. This draws on the theory of Granger causality which uses the coefficients of a bilinear model to quantify instantaneous information transfer (i.e. the relationship between signals at a fixed single time $t$) by predicting the activity of node $j$ at time, $t$, from other nodes, $i$, with a linear model (Luo et al., 2013; Mingzhou Ding, Yongheng Chen, Stephen L. Bressler, 2006).

For resting state fMRI, we want to derive a measure of functional connectivity between every set of nodes, resulting in a functional connectivity (FC) matrix. Our procedure using the covariate weights from the predictive model populates one row of the FC matrix at a time. If we repeat the process for each region, we fill the entire FC matrix by fitting $N$ models. The choice of model $M$ determines what associations we can detect between regions from the predicted covariate weights, which enables granular modeling control compared to previous attempts that use only one model (Murugesan et al., 2020). In this work we allow $M$ to be any of the following models: 1) the extremely random trees model (ERT), 2) nonlinear radial basis function kernel support-vector machine regressor (SVM), 3) Extreme Gradient Boosting forest models (XGB). The ERT was chosen because it produces high performance across a wide domain of machine learning applications (Feczko et al., 2018; Mellema et al., 2022). The SVM was chosen because it is a high-performing machine learning model which has a more directly interpretable and explicit weights than the ERT (Arora et al., 2018; Deshpande et al., 2010; Mellema et al., 2022). The XGB was chosen because it tends to have higher performance than the ERT, and handles multicollinearity from repeated data sub-sampling, which we hypothesize will better handle correlated regional information than the ERT.

For each proposed model, we use the following model fitting approach. First, the mean timeseries per region is standardized with a mean of 0 and unit variance. Then, a model is fit to predict regional activity at node $j$ at every time $t$ from other all other nodes $i$ at each time $t$. Then, a measure of feature weight or importance is extracted from the model for each nodal covariate $i$. We repeat this for each node $j$ to fully populate an asymmetric FC matrix. The asymmetric matrix is then symmetrized by averaging itself with its transpose. Feature importance is calculated from the Gini importance for the ERT, the covariate weight for the SVM, and the Gini importance weighted by number of samples routed through the decision node for the XGBoost model. The XGBoost model was fit with a group-level hyperparameter search and the ERT and SVM models were found not to benefit from this search and their default parameters were used. These hyperparameter searches were done on HCP data NOT used in training, validation, or testing. For additional model fitting details see **Supplemental section 9.1.3**. In order to evaluate the relative benefits of each proposed *ML.FC* measure, we test each FC measure's reproducibility and evaluate its predictive power by using it to infer 3 individual traits of interest (see **Section 2.4**).

**Algorithm 1:** GC Algorithm. This algorithm describes the steps by which one calculates an effective connectivity matrix $\boldsymbol{E}$ from a neural timeseries $\boldsymbol{X}$ using a standard Granger-causal approach. $\boldsymbol{X}_t^i$ = neural activity matrix of size $i$ by $t$, where $i \in [1, N]$ and where $t \in [1, T]$. $T$= number of timepoints. $\tau$=max lag. $\boldsymbol{f}$=timeseries predictor function. $j$=secondary indexer from 1 to $N$. $\boldsymbol{E}_{i,j}$=effective connectivity matrix of size $NxN$, indexed by $i$ and $j$. $\boldsymbol{r}$=reduced timeseries predictor function without region $j$. $\sigma$=standard deviation.

|  |  |
|---|---|
| **Inputs:** $\mathbf{X}_t^i, (i \in [1, N], t \in [1, T])$ ,$\tau$, $\mathbf{f}$; | Input timeseries with number of regions $N$, number of timepoints $T$, max lag $\tau$, timeseries predictor $\mathbf{f}$; |
| **Output:** $\mathbf{E}$; | Output effective connectivity matrix $E$; |
| **for** $i := 1$ *to* $N$ **do** *Full model fit* | For the initial regional timeseries $\mathbf{X}$ with $N$ regions fit a full model including region $j$; |
| $\quad \mathbf{X}_t^i = \mathbf{f}(\mathbf{X}_{t-1}, \mathbf{X}_{t-2}, ..., \mathbf{X}_{t-\tau})$; | Fit $\mathbf{f}$ predicting activity $\mathbf{X}$ at time $t$ and node $i$ from times $t-1, ... t-\tau$; |
| $\quad$ **for** $j := 1$ *to* $N$ **do** *Reduced model fit* | Fit a reduced model without region $j$; |
| $\quad\quad \mathbf{X}' = \mathbf{X}_{j \setminus N}$; | Drop column $j$ from timeseries $\mathbf{X}$; |
| $\quad\quad \mathbf{X}'_t = \mathbf{r}(\mathbf{X}'_{t-1}, \mathbf{X}'_{t-2}, ..., \mathbf{X}'_{t-\tau})$; | Fit $\mathbf{r}$ predicting activity $\mathbf{X}$ at time $t$ and node $i$ from times $t-1, ... t-\tau$ without node $j$; |
| $\quad\quad \mathbf{E}_{i,j} = log(\sigma(\mathbf{f}_{error}) / \sigma(\mathbf{r}_{error}))$; | The EC score between $i$ and $j$ equals the log of the ratio of the standard deviation of the residuals of the full and reduced model; |
| $\quad$ **end** | |
| **end** | |

### 2.1.2 Background of effective connectivity

In addition to functional connectivity, brain connectivity can be quantified with measures of time-delayed information transfer, which we denote as effective connectivity measures. Effective connectivity can be quantified in numerous ways: multivariate Granger-causal (GC) scores (Abidin et al., 2019; Chockanathan et al., 2019; Spencer et al., 2018), bilinear Granger-causal modeling (Luo et al., 2013), and other measures of directed neural influence (Bielczyk et al., 2019). This paper builds new measures from the mathematical foundation of Granger causal modeling. Granger-causal measures define a directed edge by quantifying how the past history of activity signal *B* from a particular brain region informs the future activity of signal *A*, from another brain region. In neuroimaging, signal *B* is said to be Granger causal of signal *A* if a model to predict the future of *A* given all past information from all regions' signals including *B* is more accurate than a model that doesn't include *B*. The degree of causality is called the GC score (Granger, 1969). To generate a Granger causal effective connectivity matrix, the Granger score between the regional time courses from each pair of regions is calculated using the GC Algorithm (**Algorithm 1**). A full model, *f*, is fit to predict activity in region *i* at time *t* from the past history of all regions. Then, a reduced model *f'* is fit to predict the same activity at time *t* from the past history of all regions except *j*. The EC score is the log of the ratio of the standard deviation of the residuals of the full and reduced models. By using a linear model *f*, a baseline measure of effective connectivity can be calculated. The linear models with which we calculate the GC score include: an unpenalized multivariate autoregressive (MVAR) model denoted $MV.GC$, an elastic multivariate autoregressive model with a small $L_1$ and $L_2$ penalty ($L_1$ = $L_2$ = $\lambda$ = 0.1) denoted $MV.GC_{E:\lambda=0.1}$, and an elastic multivariate autoregressive model with a large L1 and L2 penalty ($\lambda$=10) denoted $MV.GC_{E:\lambda=10}$. These regularization amounts were chosen empirically to be representative of strong and weak regularization. The timeseries is tested for significant autoregression with the Augmented Dickey Fuller test and any significant autoregression is removed prior to model fitting. Lag values of 1-5 times repetition time (TR) were tested and the model using the lag with the lowest Akaike information criterion (AIC) was selected independently for each regional model.

**Algorithm 2:** MLEC algorithm. This algorithm describes the steps by which one calculates an effective connectivity matrix $\boldsymbol{E}$ from a neural timeseries $\boldsymbol{X}$ using a standard Machine-learning effective connectivity. $\boldsymbol{X}_t^i$ = neural activity matrix of size $i$ by $t$, where $i \in [1, N]$ and where $t \in [1, T]$. $T$= number of timepoints. $\tau$=max lag. $\boldsymbol{f}$=timeseries predictor function. $G$=operator which calculates the importance score of a given model, this is the Gini impurity in a tree-based model or weight in an SVM-based model. $j$=secondary indexer from 1 to $N$. $C$=total number of unique samples in $\boldsymbol{X}$. $P$=proportion of data routed through split. $p$=probability of data routed to a split. $w$=classification vector of SVM. $\eta$=learned regularization parameter, $\theta$=learned transform of $x$, b=SVM bias, $\boldsymbol{E}_{i,j}$=effective connectivity matrix of size $NxN$, indexed by $i$ and $j$.

| | |
|---|---|
| **Inputs:** $\mathbf{X}_t^i, (i \in [1, N], t \in [1, T]), \tau, \mathbf{f}$; | Input timeserieswith number of regions $N$, number of timepoints $T$, max lag $\tau$, timeseries predictor $\mathbf{f}$; |
| **Output:** $\mathbf{E}$; | Output effective connectivity matrix $E$; |
| **for** $i := 1$ *to* $N$ **do** *Full model fit* | For an initial regional timeseries $\mathbf{X}$ with $N$ regions fit a full model including region $i$; |
| $\mathbf{X}_t^i = \mathbf{f}(\mathbf{X}_{t-1}, \mathbf{X}_{t-2}, ..., \mathbf{X}_{t-\tau})$; | Fit a machine learning model $\mathbf{f}$ predicting activity $\mathbf{X}$ at region $i$ at time $t$ from times $t-1, ... t-\tau$; |
| **if** $f$ = *ERT or XGB* **then** | If an XGBoost or extremely random trees predictor; |
| $G(f)_j = \sum_{n=1}^{C} P(p(n) * 1 - p(n))$; | Importance score between $i$ and $j$ equals the Gini impurity for feature $j$ with probability of the data being routed down a split $p$, proportion of data routed to that split $P$, set of all nodes that use feature $j$ of $C$; |
| **else if** $f$ = *SVM* **then** | If a support vector predictor; |
| $G(f) = w \mid (min\|\|w\|\|^2 + C\sum(\eta_i)) : (y_i(w \cdot \theta(x_i) + b) \geq 1 - \eta_i, \eta_i \geq 0)$; | Importance score is the weight given feature $j$ by classification vector $w$ given the support vector optimization with regularization terms $\eta$, $\theta$, $b$; |
| $\mathbf{E}_i = G(f)$; | The EC score between $i$ and $j$ is the feature importance for $j$ of the model $f$ predicting $i$; |
| **end** | |

Next, we will build upon this classical GC foundation in two ways: by replacing the GC's linear multivariate autoregressive (MVAR) model with a nonlinear multivariate[2] machine learning model (explained in Section 2.1.3), and by fitting the GC models in a lower dimensional space with a dimensionality reduction that also enforces biological constraints from structural connectivity (explained in Section 2.1.4).

### 2.1.3 Proposed machine learning-based effective connectivity ($\boldsymbol{ML.EC}$) measures

We propose a novel EC measure using machine learning model coefficients as in **Section 2.1.1**, but with a time-delay lag, τ, included. The machine learning model predicts future timesteps given up to τ past timesteps and identifies the important learned features of that model, (**Algorithm 2**). Compared to GC, our proposed measure captures directed influence with only one model fit per region analyzed, and thereby scales as $O(N)$ rather than $O(N^2)$ where $N$ is the number of regions. Furthermore, the machine learning approaches capture nonlinear interactions which the standard GC approaches do not. We denote this Machine learning effective connectivity approach $\boldsymbol{ML.EC}$. We test both $ML.EC$ with an extremely random trees internal predictor (denoted $ML.EC_{ERT}$ ) and a support vector regressor with a radial basis function kernel, denoted $ML.EC_{SVM}$.

### 2.1.4 Proposed structurally-projected effective connectivity ($\boldsymbol{SP.GC}$) measures

We also propose another novel EC measure which projects the Granger causal models into a lower dimensional space informed by a prior from diffusion MRI. This is a soft constraint which regularizes the EC measure to have at least some agreement with known physical pathways of communication (Allen and Weylandt, 2019; Dillon et al., 2017; Huang and Ding, 2016; Maglanoc et al., 2020; Manning et al., 2018). We denote this Structurally Projected Granger Causality approach $\boldsymbol{SP.GC}$, (see **Algorithm 3**). This approach projects the timeseries into a lower dimensional representation and calculates a full and a reduced model in the lower dimensional space before projecting the predicted activity back into the original space and calculating the error in full versus reduced models. This encourages but does not force, low dimensional timeseries components to lie along known structural networks. This approach also incorporates a sparsity constraint from sparse PCA. Sparse PCA minimizes the number of nonzero terms in each principal component while maximizing the variance explained by the components (Zou and Xue, 2018). This sparsity prior and the prior from the structural connectivity matrix derived from diffusion MRI encourages these components to robustly represent a physically connected sub-network of the brain.

[2] We use the term multivariate to indicate a multivariable model with multiple independent variables.

**Algorithm 3:** SPGC algorithm. This algorithm describes the steps by which one calculates an effective connectivity matrix $\boldsymbol{E}$ from a neural timeseries $\boldsymbol{X}$ using a standard Granger-causal approach. $\boldsymbol{X}_t^i$ = neural activity matrix of size $i$ by $t$, where $i \in [1, N]$ and where $t \in [1, T]$. $T$= number of timepoints. $\Phi$=transformation matrix from learned lower-dimensional transofrm. $\tau$=max lag. $\theta$=low-dimensional representation of neural activity matrix $\boldsymbol{X}$. $\boldsymbol{f}$=timeseries predictor function. $j$=secondary indexer from 1 to $N$. $\boldsymbol{E}_{i,j}$=effective connectivity matrix of size $NxN$, indexed by $i$ and $j$. $\boldsymbol{r}$=reduced timeseries predictor function without region $j$. $\sigma$=standard deviation.

| **Inputs:** $\mathbf{X}_t^i, (i \in [1, N], t \in [1, T]), \Phi, \tau, \mathbf{f}$; | Input timeseries with number of regions $N$, number of timepoints $T$, transformation matrix $\Phi$ from prior-informed sparse PCA (Equation 2), max lag $\tau$, timeseries predictor $\mathbf{f}$; |
|---|---|
| **Output:** Effective connectivity matrix $\mathbf{E}$; | Output effective connectivity matrix $E$; |
| **for** $i := 1$ *to* $N$ **do** *Full model fit* | For the initial low dimensional regional timeseries $\theta$ fit a full model including region $j$ in the low dimensional projection; |
| $\theta = \mathbf{X} \cdot \Phi$; | Project the initial regional timeseries $\mathbf{X}$ to structurally constrained subspace $\theta$ using transformation matrix $\Phi$ (Equation 2); |
| $\mathbf{X}_t^i = \mathbf{f}(\theta_{t-1}, \theta_{t-2}, ..., \theta_{t-\tau}) \cdot \Phi^T$; | Fit $\mathbf{f}$ predicting activity $\mathbf{X}$ at time $t$ and node $i$ from $\theta$ at times $t-1, ... t-\tau$; |
| **for** $j := 1$ *to* $N$ **do** *Reduced model fit* | Fit a reduced model without region $j$; |
| $\theta' = \mathbf{X} \cdot \Phi_{j \setminus N}$; | Project into structurally constrained low dimensional space $\theta'$ with $\Phi_{j \setminus N}$, the transformation matrix with column $j$ removed; |
| $\mathbf{X}_t^i = \mathbf{r}(\theta'_{t-1}, \theta'_{t-2}, ..., \theta'_{t-\tau}) \cdot \Phi_{j \setminus N}^T$; | Fit $\mathbf{r}$ predicting activity $\mathbf{X}$ at time $t$ and node $i$ from $\theta'$ at times $t-1, ... t-\tau$; |
| $\mathbf{E}_{i,j} = log(\sigma(\mathbf{f}_{error}) / \sigma(\mathbf{r}_{error}))$; | The EC score between $i$ and $j$ equals the log of the ratio of the standard deviation of the residuals of the full and reduced model; |
| **end** | |
| **end** | |

We implement a structural connectivity constraint which encourages a more faithful interpretation the underlying brain functional from fMRI as true functional connectivity lies sparsely along physical connections (Allen and Weylandt, 2019; Dillon et al., 2017; Huang and Ding, 2016; Maglanoc et al., 2020; Manning et al., 2018). The physical connectivity prior comes from a structural connectivity (SC) matrix derived from diffusion MRI. The structural connectivity matrix is calculated from the average normalized tractogram from all 1065 subjects in the human connectome project (HCP) computed in (Yeh et al., 2018).

We use a population-level prior so that *it may be applied even when diffusion MRI is unavailable for every individual in a study*. The strength of the structural connectivity between each pair of brain regions is computed from the number of tractography streamlines passing through each region of interest in this normalized, ensemble atlas. Then the log of the total number of streamlines between each region is taken to be used as the prior. See **Supplemental section 9.1.4** for further details. There are a plethora of both direct and indirect connections in the brain, and both direct and indirect connections are captured through the streamline-derived prior. Each streamline can represent multisynaptic or monosynaptic fibers. In formulating a prior from SC, we encourage activity to lie along the streamline populations. This is an intuitive, interpretable, and logical constraint to add to EC. If there is a strong structural connection between two regions or voxels, it is *more likely* to have a substantive connection between them. Furthermore, physiologically, regions that are not connected *should not* communicate without traveling through intermediate regions. Imposing the constraint of a prior along which we hypothesize communication should lie helps to (though does not completely) prevent inferring erroneous connections. This prior in particular is relevant for effective connectivity approaches, where we are attempting to untangle directionality and indirect versus direct communication pathways. So, connections mediated mostly by intermediate nodes, such as an indirect connection from region A to region C through intermediate region B, are explicitly modeled rather than a connection from region A to C being inferred directly. While the concept of using structural connectivity from tractography to constrain fMRI interpretation has been used to interpret fMRI before (Huang and Ding, 2016; Maglanoc et al., 2020), the combination with dimensionality reduction is novel to this work.

We use a formulation of prior-constrained sparse PCA to incorporate our SC prior in the timeseries dimensionality reduction. The objective function for this constrained sparse PCA shown in **Equation 2**; which we have adapted from Dhillon et al. (Dhillon et al., 2014) to include the tractography prior:

$$\vec{v_i}^* = \underset{v_i, ||v_i||=1, v_i^T, v_j=0, i \neq j, v_i \geq 0}{argmax} (\vec{v_i}^T (\mathbf{C} + \theta \cdot \mathbf{P}_i^T \mathbf{P}_i) \vec{v_i} - \lambda \cdot ||\vec{v_i}||_1), \quad \mathbf{P}_{i,j} = log(S_{i \cap j}) \tag{2}$$

The first term in the objective function enforced data fidelity. It depends on the covariance $\boldsymbol{C}$ ($N_{ROI} \times N_{ROI}$) between mean pairwise mean regional timeseries but is regularized with a structural prior $\boldsymbol{P}$ with an initial belief $\theta$. The second term, $\lambda \cdot \|\vec{v_i}\|_1$ , imposes the $L_1$ sparsity with weight $\lambda$. The prior $\boldsymbol{P}$ is a reformulation of the SC matrix into a matrix where the rows correspond to the individual regions and the columns correspond to a larger network each region can be grouped into. See **Supplemental section 9.1.4** for further details.

This approach has three advantages over the standard Granger causal measure. *First*, it incorporates prior information to regularize fMRI interpretation. *Second*, it reduces the number of measure parameters that must be tuned. In *SP.GC*, only the maximum lag needs be selected for the $SP.GC$ method as the prior belief weight ($\theta$ in **Equation 2**) is fixed at 1 for all experiments, giving equal weighting to the calculated covariance and the prior. GC with an elastic net requires selection of: 1) an L1/L2 ratio, 2) penalty weight, and 3) testing of multiple lags up to and including the maximum lag. In practice, for a given maximum lag, only 1 model needs to be fit for $SP.GC$, while fitting the standard GC requires dozens of cross-validated models to be fit to properly optimize the lag and regularization parameters. A sufficiently large lag selected before dimensionality reduction appropriately weights lag values through the dimensionality reduction itself (DSouza et al., 2017). *Third*, $SP.GC$ model fitting is faster than standard GC, as there are fewer variables as the number of components is much less than the number of regions.

In this study, the Schaefer functional atlas (Schaefer et al., 2018) is used with the cerebellum and striatum added from the AAL atlas (Rolls et al., 2020). The Schaefer atlas has the advantage that each ROI is assigned a corresponding coarse and fine RSN label facilitating the construction of our RSN structural connectivity prior. We generate two SC network priors, a *coarse* one with 18 regions (7 RSNs, subcortical structures, and the left and right hemisphere cerebellar gray matter), as well as a *fine* prior with 38 regions (17 RSNs, subcortical structures, and the left and right hemisphere cerebellar gray matter). The coarse SC prior encourages functional connections that capture left/right hemisphere connections at a whole RSN level, while the fine SC prior encourages the projection of whole brain functional activity to smaller subsections of structurally connected sub-elements of the larger RSNs. We denote the $SP.GC$ approach using the coarser 7 RSN prior $SP.GC_{c:7}$ and the $SP.GC$ approach using the finer 17 RSN prior $SP.GC_{f:17}$.

As a baseline of comparison, we also choose to use a PCA projection to a number of components preserving 95% of the variance in the timeseries, analogous to previous work (Abidin et al., 2019; Chockanathan et al., 2019; Luo et al., 2013). The baseline measure does not impose the structural connectivity prior. This measure can be computed with **Algorithm 3,** using a PCA projection rather than a structurally-constrained projection. We denote this PCA-projected low dimensional Granger scored measure $PC.GC$.

### 2.2 Materials

This work uses fMRI data from the Human Connectome Project (HCP) (van Essen et al., 2012) to evaluate the proposed connectivity measures. We use the 4 scans of each participant, including on one day: (i) a left-to-right phase-encoded fMRI acquisition and (ii) a right-to-left phase-encoded fMRI acquisition, and on a subsequent day: a repeated (iii) left-to-right and (iv) a right-to left acquisition. The measures are evaluated for their ability to produce a consistent connectivity matrix across the repeat scans for each subject.

From the HPC database, 805 subjects have the full complement of 4 repeat scans and demographic information. From these we excluded subjects with substance use (including alcohol and tobacco), as these are known to confound the reproducibility of longitudinal connectivity. Of the remaining 517 subjects, we selected 100 subjects which were demographically diverse and had the least head motion defined by mean framewise displacement between fMRI frames. This minimizes motion confounds which can influence connectivity measures with correlated non-neural signal (Noble et al., 2019; Satterthwaite et al., 2019). Subjects were selected to match the demographics of the 2010 USA census data (subject distribution shown in **Figure 1B**) and the CONSORT diagram of data selection is shown in **Figure 1A**. I.E., as the 2010 census distribution was 61.5% white, 17.6% Hispanic/Latino, 12.3% Black, 8.6% Other, 62 white, 18 Hispanic, 12 Black, and 8 Other subjects were chosen, with an even male/female split. A plot of the mean framewise displacement of the selected subset of 100 subjects versus the 517 initial subjects (**Figure 1B**), shows the motion level of the chosen set compared to the remainder of HCP under consideration. This selected subset of subjects is later further split for cross-validation of reproducibility and for nested cross-validation for predicting of target values (with multiple train, validation, and test splits in the nested case).

The selected data was processed with the standard HCP minimal preprocessing pipeline (Glasser et al., 2013). Mean regional timeseries were then extracted with the Schaefer atlas with 100 anatomical regions with additional subcortical regions included (Schaefer et al., 2018). The parcellation of 100 was chosen over the 200 or 400 region parcellation to decrease computation time while remaining sufficiently large to test our approaches. The Schaefer atlas is a functional atlas whose regions are defined through the clustering of functional activity in fMRI (Schaefer et al., 2018), and this functional atlas was chosen because: 1) a functional atlas tends to capture better functional variability than a purely anatomical atlas (Mellema et al., 2022), and 2) the Schaefer atlas groups regions into resting-state networks (RSN), which facilitates inter and intra RSN partitioning and analysis. The cerebellum and striatum from the AAL atlas (Rolls et al., 2020) were included as well because both structures contain signals of diagnostic importance and are often overlooked in prior analyses (Stoodley et al., 2012).

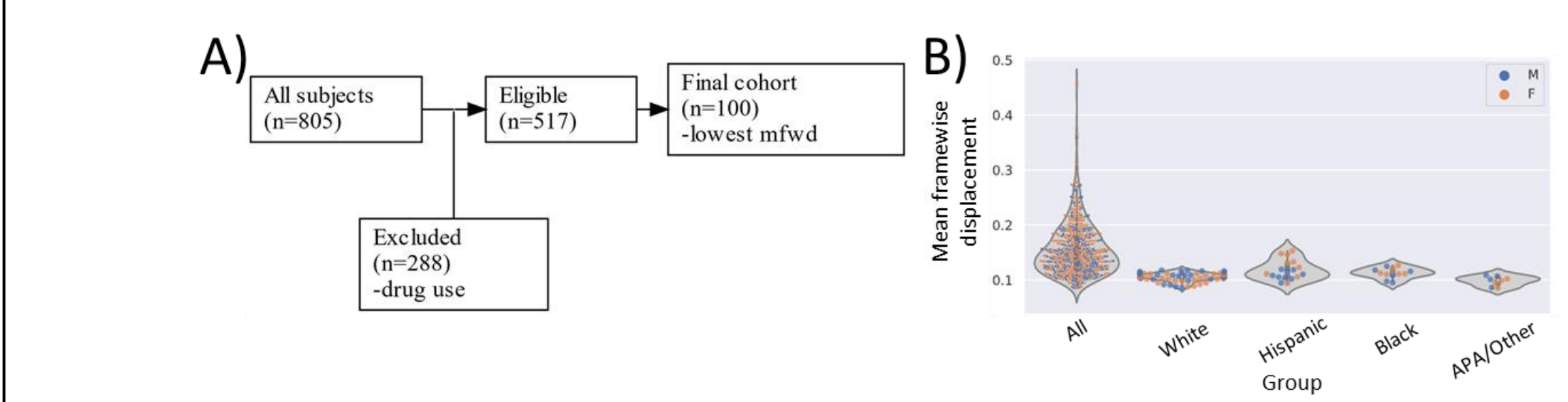

**Figure 1: Subject selection.** A) CONSORT diagram of data selection. Subjects were filtered by completeness of data, concomitant drug use, and then sorted by lowest mean framewise displacement. A demographically representative set of the 100 lowest motion subjects was chosen for the final analysis. B) Efficacy of HCP partitioning. The distribution of mean framewise displacement for all subjects is shown versus each of the selected demographic subsets. Male subjects are indicated with a blue dot while female subjects are indicated with an orange dot.

### 2.3 Experiment 1: Comparison of reproducibility

The reproducibility of the connectivity measures was quantified using five different metrics. These reproducibility metrics include linear, nonlinear, and clustering metrics. The linear metrics included: (1) the average root mean squared difference of each element in the connectivity matrix (after z-score normalizing the elements) across runs of the same subject, and (2) the Pearson's Correlation, *r,* between the connectivity elements (edges) of any 2 pairs of scans of the same subject, averaged over all pairs of scans. The nonlinear metrics of reproducibility included: (1) the average cosine similarity between all edges of any 2 pairs of scans of the same subject, and (2) the two-way random, single score intraclass correlation coefficient (ICC(2,1)) edgewise (Noble et al., 2019) between any 2 scans of the same subject. Additionally, a clustering score for each EC or FC measure was calculated. This clustering score was the Davies-Bouldin (DB) index (Bezdek and Pal, 1998). This index quantifies how well each subject is separated from all the subjects after projection to a low dimensional space and a higher value indicates greater separation (Finn et al., 2015). A higher DB score indicates that a more subject-specific fingerprint was identified, capturing aspects of connectivity unique to that subject. Finally, as an additional post-hoc analysis we evaluated how well each measure performed when given less and less of the timeseries data duration. This tested each measure's ability to maintain high reproducibility using a fraction of the timeseries, which could enable shorter acquisitions facilitating future fMRI studies.

### 2.4 Experiment 2: Comparison of predictability of individuals' traits

In addition to reproducibility metrics, using the subset of EC and FC measures that had the highest reproducibility we tested how well the connectivity measures could predict 3 categories of targets: a physiological trait, cognitive trait, and a combined physiological and cognitive trait. The physiologic trait chosen from the HCP dataset was *mean arterial blood pressure*, the cognitive trait was *fluid intelligence* as measured with the Pennsylvania matrix reasoning test, and the combined physiologic and cognitive trait was the *stress and adversity inventory*. The combination of physiologic, cognitive, and combined traits was chosen to be more representative of possible real-world targets than a random selection of predictable values. These targets each are a single scalar value per subject averaged over their multiple visits if obtained more than once. I.E., for the first subject's four scans, there is one target scalar the models are trying to predict from each of the connectivity 'fingerprints'. Cross-validation splitting was performed at a subject level for all these analyses, at no point was a subject's scans split between training and validation.

### 2.4.1 Univariate analysis of effect size

To evaluate how predictive the connectivity measures are, first we performed a univariate analysis. The effect size per edge was measured with Cohen's d and compared across the measures with high reproducibility, as measured by Pearson's *r* and measured with the individual subject clustering score in order to compare predictive information present in the calculated connectivity. An ideal measure will have high reproducibility and high effect size across a variety of predictive tasks.

### 2.4.2 Multivariate analysis

Univariate analysis does not suffice to show that a connectivity measure is apt to yield accurate *multivariate* predictions. Therefore, we also selected *significant edges* (univariate significance $p \leq 0.05$) as candidate covariates to form multivariate predictive models for each connectivity measure. To reduce collinearity, pairs of such edges with covariance greater than 0.75 were identified and the edge with higher effect size was retained. The surviving set of edges was used to predict the targets as in experiment 1 with an elastic net predictor and the predictive power was measured. A 10x5 cross-validation approach was applied. In the 10-fold cross validation, the data was stratified by the target measure and 10% of subjects were set aside to test each iteration. The inner 5-fold cross validation performed hyperparameter optimization where the elastic net's L1 /L2 ratio was tuned with a grid search using ratios from the set {0.1, 0.5, 0.7, 0.9, 0.95, 0.99, 1.0}. A nested cross-validation approach has been shown to better represent real-world performance (Cawley and Talbot, 2010). The regularization weight, α, was optimized through coordinate descent, and the model with the lowest mean squared error across the inner cross validation folds was evaluated on the 10% test set from the outer fold.

The entire procedure was itself repeated 10 times using Monte Carlo iteration as this has been shown to increase the stability of the estimated prediction (Wu et al., 2020). The coefficient of determination, $R^2$, of the internally fitted model on the held-out data and averaged over outer Monte Carlo iterations was recorded. A paired one-tailed t-test testing the null hypothesis that the highest performing model's $R^2$ was not greater than other models' $R^2$ was performed. The Bonferroni corrected p-value was reported. Bonferroni correction was chosen to minimize type 1 error with a criterion more stringent than FDR correction for answering the specific question of which predictor produces the highest $R^2$.

### 2.4.3 Multi-input, multivariate analysis

An additional post-hoc analysis was performed to further measure the complementarity of the connectivity measures from the trait predictions described in **Section 2.4.2**. This secondary analysis determines which connectivity measures contain complementary information and compares models built with a combination of complimentary measures to the models built on the original separate measures. To determine measure complementarity, a linear mixed effects (LME) model was fit to predict a participants' trait by combining the predictions from multiple models each using a single separate connectivity measure. For example, the predicted mean arterial pressure predicted from *partial correlation* was concatenated with the predicted arterial pressure from the *other FC connectivity measures* into a vector, and that vector of predictions was used to predict mean arterial pressure. If partial correlation contains trait predictive information that $ML.FC_{ERT}$ does not and vice-versa, we would expect the LME model to give significant weight to predictions generated from both partial correlation and $ML.FC_{ERT}$. If one of the measures contains only a subset of the information contained by another, the measure with greater predictive information will be assigned a large weight, whilst the other measure will be assigned a small weight close to zero. The LME models were fit with a subject-specific intercept and group level slope per **Equation 3**:

$$y_j = \mu_{0,j} + \beta_1 x_{1,j} + \beta_2 x_{2,j} + \cdots + \beta_n x_{n,j} + \epsilon_j \tag{3}$$

where *j* indexes the subject and *n* is the number of measures combined from **Section 2.4.2**. The predicted trait for a subject, $y_j$, is a function of the subject-specific intercept ($\mu_{0,j}$) and the weights ($\beta_i$) on the prediction from each of the $n$ measures. Each predictor ($x_{n,j}$) is the predicted value of subject $j$'s trait (e.g., predicted mean arterial pressure, stress, or fluid intelligence) from the each of the $n$ elastic models trained from a single measure (e.g., partial correlation). The fitted LME coefficients with $p \leq 0.05$ and magnitude greater than 10% of the maximum coefficient magnitude were considered to contain complimentary information in the predictions. This produced a subset of complimentary connectivity measures that could then be used in another multi-input, multivariate elastic net model. This secondary model with complimentary measures was also fit with 10x5 cross-validation and compared to the original models, to test the benefit of combining the complimentary features.

# 3 Results

## 3.1 Experiment 1: Comparison of reproducibility

### 3.1.1 Comparison of the reproducibility of functional connectivity

The reproducibility of six functional connectivity metrics: 1) Pearson's Correlation based connectivity, 2) partial correlation connectivity, 3) spectral GC functional connectivity (denoted $SG.FC$ ), and $ML.FC$ based connectivity 4) with XGB, 5) with extremely random trees, and 6) with a radial basis function support vector regressor was quantified with five metrics. These metrics included: Pearson's *r*, Root Mean Square Error (RMSE), intraclass correlation coefficient (ICC), cosine similarity, and the *ease of separability* (DB clustering score) as described in **Section 2.3**. The results from these comparisons are shown in **Table 1** with the 5 right most columns showing the mean reproducibility and 95% confidence interval. The best reproducibility in each column is in **boldface**. The significant differences (FDR corrected *p*-value < 0.05) between the top result and the other entries within a column are denoted with an asterisk, $*$. FDR Correction was chosen over familywise type 1 error rate control as we have higher tolerance for type 1 errors and lower tolerance for type 2 errors in this relatively broad comparison using multiple measures. The proposed metrics are distinguished with a grey background. Of all the FC metrics, the proposed $ML.FC_{XGB}$ connectivity had the highest Pearson's *r* and cosine similarity, as well as the lowest RMSE. Partial correlation connectivity had the best (lowest) DB index while, correlation-based connectivity had a higher ICC than any other metric; however, this metric is less sparse than the other metrics which could artificially inflate this measure for correlation. Though RMSE is a suboptimal metric of reproducibility across multiple metric types due to the different distributions and sparsity of the FC metrics, it is shown in order to facilitate comparisons to the literature. Pearson's *r* which is less susceptible to scale variations and sparsity, and the DB clustering score, which captures subject identifiability, were selected as measures for further analysis. A comparison of the six FC methods along these two metrics is shown in **Figure 2.** Under the X axis, the mean connectivity matrix is shown which is computed across all subject scans for each method. The ordering of the anatomy and RSNs across the columns and rows within each matrix is detailed in **Supplementary Figure S1**. The proposed $ML.FC_{XGB}$ with XGB connectivity metric (right) had the best reproducibility by most measures (**bold** in **Table 1**) including Pearson's *r*, while partial correlation had the second best for most measures except DB score, for which it had the best. This suggests conditioning the connectivity between two nodes on the activity of all other nodes is critical for reproducibility.

**Table 1: Comparison of the reproducibility of the six FC methods including proposed (gray shaded) and traditional (nonshaded) functional connectivity measures.** The section of this article where the measure is described is indicated in the Section column. The best performing methods in each metric are shown in **boldface**. Statistically significant differences from the best performing method after FDR correction at 5% are indicated with *. $Corr$=correlation, $PCorr$=partial correlation, $SG.FC$=spectral Granger causality functional connectivity, $ML.FC_{ERT}$=machine learning functional connectivity using an extremely random trees predictor, $ML.FC_{SVM}$=machine learning functional connectivity using a support vector regressor, $ML.FC_{XGB}$=machine learning functional connectivity using an XGB model.

| FC measure ( =ours) | Section | Linear | | Nonlinear | | Clustering |
|---|---|---|---|---|---|---|
| | | Pearson's r | RMSE | ICC(2,1) | Cosine similarity | Clustering DB Score |
| $Corr$ | **1** | $0.40 \pm 0.00^*$ | $0.70 \pm 0.02^*$ | $\mathbf{0.43 \pm 0.0}$ | $0.75 \pm 0.01$ | $10.5 \pm 1.0^*$ |
| $PCorr$ | **1** | $0.44 \pm 0.00^*$ | $0.49 \pm 0.00^*$ | $0.07 \pm 0.00^*$ | $0.88 \pm 0.00^*$ | $\mathbf{0.23 \pm 0.01}$ |
| $SG.FC$ | **1** | $0.31 \pm 0.01^*$ | $0.92 \pm 0.02^*$ | $0.29 \pm 0.00^*$ | $0.57 \pm 0.02^*$ | $13.2 \pm 1.0^*$ |
| $ML.FC_{ERT}$ | **2.1.1** | $0.42 \pm 0.00^*$ | $0.56 \pm 0.01^*$ | $0.31 \pm 0.00^*$ | $0.84 \pm 0.01^*$ | $5.55 \pm 0.55^*$ |
| $ML.FC_{SVM}$ | **2.1.1** | $0.35 \pm 0.00^*$ | $0.79 \pm 0.01^*$ | $0.27 \pm 0.00^*$ | $0.68 \pm 0.01^*$ | $8.00 \pm 0.97^*$ |
| $ML.FC_{XGB}$ | **2.1.1** | $\mathbf{0.46 \pm 0.00}$ | $\mathbf{0.40 \pm 0.00}$ | $0.06 \pm 0.00^*$ | $\mathbf{0.93 \pm 0.00^*}$ | $3.14 \pm 0.30^*$ |

### 3.1.2 Comparison of the reproducibility of effective connectivity

The *effective* connectivity measures were evaluated with Pearson's *r*, RMSE, ICC, cosine similarity, and the ease of separability via the DB score, as described in **Section 2.3.** The results from these comparisons are shown in **Table 2**. Significant differences are denoted with an asterisk and the most reproducible measure in each column is **boldfaced**. The proposed metrics are distinguished with a grey background. We note that the ICC metric for measuring the reproducibility of the GC connectivity measures is influenced by the sparsity of the connectivity measure. The proposed $ML.EC_{ERT}$ connectivity with extremely random trees predictor outperformed all other traditional and proposed methods across *all* metrics except clustering, where it provided a respectable performance close to the median among the tested methods. The proposed $ML.EC_{ERT}$ connectivity outperformed the proposed $ML.EC_{SVM}$ with SVM implementation across all reproducibility measures, suggesting superiority of the ERT based predictor for this connectivity measure.

The regularized GC connectivity measures from the elastic multivariate autoregressive model ($MV.GC_E$) performed significantly better than the unregularized Granger-causal measure ($MV.GC$). The higher elastic penalty ($\lambda = 10$) increased Pearson's *r* relative to the lower elastic penalty ($\lambda = 0.1$), and the lower elastic penalty had a superior cluster separability. The proposed structurally projected GC method, $SP.GC_{f:17}$, attained greater reproducibility than either the elastic $MV.GC_E$ or the $PG.GC$ connectivity measures in its cosine similarity, and was also better in other measures of reproducibility including Pearson's r and RMSE. To further understand the differences between the Granger-causal connectivity metrics (*MV.GC*, *PC.GC* and *SP.GC*), we quantified the stability of these measures as a function of the amount of scan time (fMRI timeseries length) used to measure the causal connectivity. Notably, $SP.GC_{f:17}$ using 50% of the initial timeseries has a higher reproducibility than the $MV.GC_E$ or $PG.GC$ methods using 100% of the timeseries.

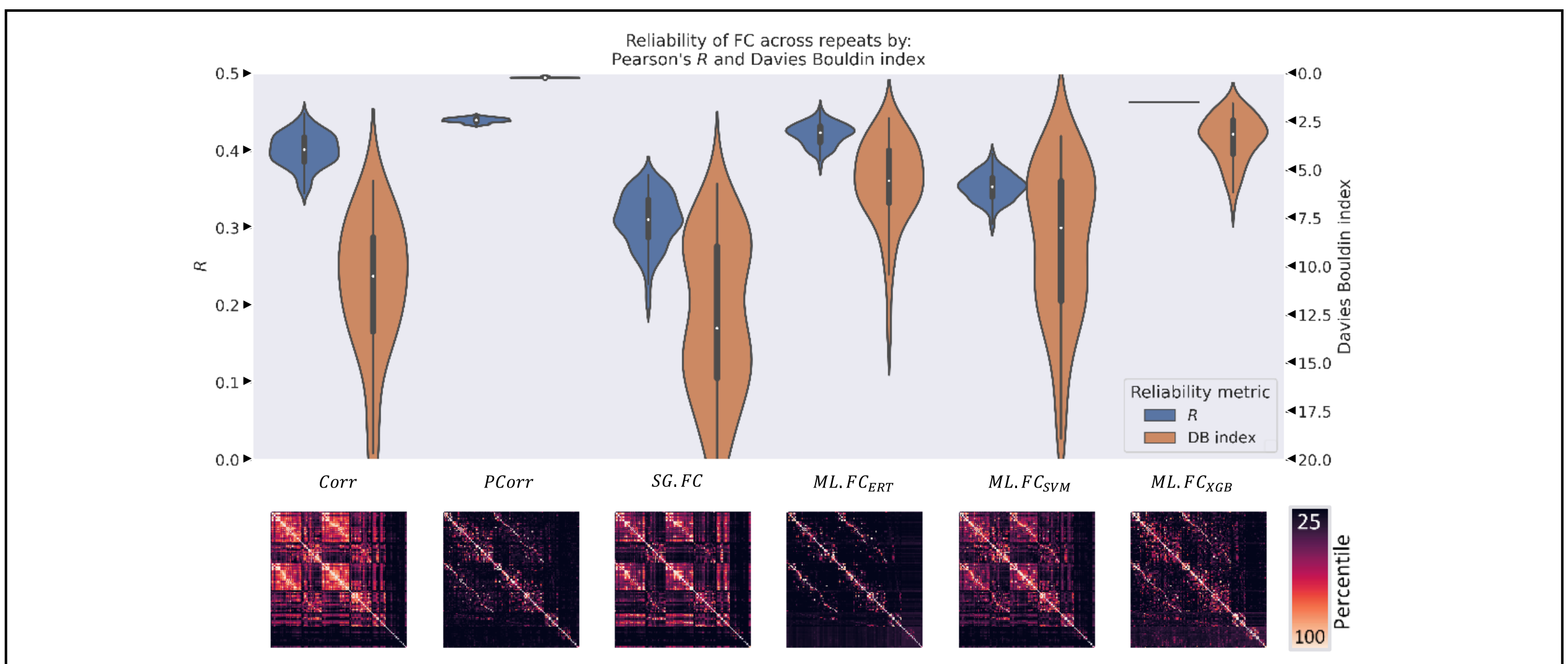

**Figure 2: Reproducibility of functional connectivity measures across repeat scans as measured by Pearson's *r* and Davies Bouldin Index**. This figure shows the distribution of reproducibility as measured by Pearson's *r* and DB clustering score for every pair of FC matrices in the four repeat-scan set per subject. Superior reproducibility for both metrics is found at the top of the graph. The mean FC matrix for each measure is displayed under the x axis.

**Table 2: Comparison of the reproducibility of the six EC methods including proposed (gray shaded) and traditional (nonshaded) functional connectivity measures.** The section where the measure is described is indicated in the Section column. The best performing methods in each metric are shown in **boldface**. Statistically significant differences from the best performing method after FDR correction at 5% are indicated with *. $MV.GC$=Granger causality measured with an unpenalized MVAR model, $MV.GC_{E:\lambda=0.1}$= Granger causality measured with an elastic MVAR model with equal $L_1$ and $L_2$ penalties and a regularization parameter of 0.1, $MV.GC_{E:\lambda=10}$ = Granger causality measured with an elastic MVAR model with equal $L_1$ and $L_2$ penalties and a regularization parameter of 10, $ML.EC_{ERT}$=Machine learning effective connectivity using an extremely random trees predictor, $ML.EC_{SVM}$ =Machine learning effective connectivity using a support vector machine regressor, $PC.GC$=low dimensional Granger causality measured with an MVAR model in PCA space, $SP.GC_{c:7}$ =structurally projected Granger causality measured with an MVAR model in the low dimensional space informed by the structual prior using 7 resting state network sub-parcelations, $SP.GC_{f:17}$ =structurally projected Granger causality measured with an MVAR model in the low dimensional space informed by the structual prior using 17 resting state network sub-parcelations.

| EC measure ( =ours) | Section | Linear | | Nonlinear | | Clustering |
|---|---|---|---|---|---|---|
| | | Pearson's r | RMSE | ICC(2,1) | Cosine similarity | Clustering DB Score |
| $MV.GC$ | **2.1.2** | $0.17 \pm 0.00^*$ | $1.15 \pm 0.01^*$ | $0.10 \pm 0.00^*$ | $0.34 \pm 0.01^*$ | $11.0 \pm 1.1^*$ |
| $MV.GC_{E:\lambda=0.1}$ | **2.1.2** | $0.29 \pm 0.00^*$ | $0.93 \pm 0.01^*$ | $0.01 \pm 0.00^*$ | $0.57 \pm 0.01^*$ | $3.97 \pm 0.39^*$ |
| $MV.GC_{E:\lambda=10}$ | **2.1.2** | $0.32 \pm 0.01^*$ | $0.85 \pm 0.02^*$ | $0.00 \pm 0.00^*$ | $0.64 \pm 0.01^*$ | $5.78 \pm 0.44^*$ |
| $ML.EC_{ERT}$ | **2.1.3** | $\mathbf{0.47 \pm 0.00}$ | $\mathbf{0.37 \pm 0.01}$ | $\mathbf{0.39 \pm 0.00}$ | $\mathbf{0.93 \pm 0.00}$ | $6.50 \pm 0.52^*$ |
| $ML.EC_{SVM}$ | **2.1.3** | $0.31 \pm 0.00^*$ | $0.90 \pm 0.01^*$ | $0.22 \pm 0.00^*$ | $0.59 \pm 0.01^*$ | $10.4 \pm 1.4^*$ |
| $PC.GC$ | **2.1.4** | $0.33 \pm 0.01^*$ | $0.82 \pm 0.02^*$ | $0.15 \pm 0.00^*$ | $0.67 \pm 0.01^*$ | $6.84 \pm 0.36^*$ |
| $SP.GC_{c:7}$ | **2.1.4** | $0.32 \pm 0.00^*$ | $0.84 \pm 0.01^*$ | $0.07 \pm 0.00^*$ | $0.65 \pm 0.01^*$ | $\mathbf{3.00 \pm 0.36}$ |
| $SP.GC_{f:17}$ | **2.1.4** | $0.36 \pm 0.00^*$ | $0.76 \pm 0.01^*$ | $0.06 \pm 0.00^*$ | $0.73 \pm 0.01^*$ | $4.32 \pm 0.39^*$ |

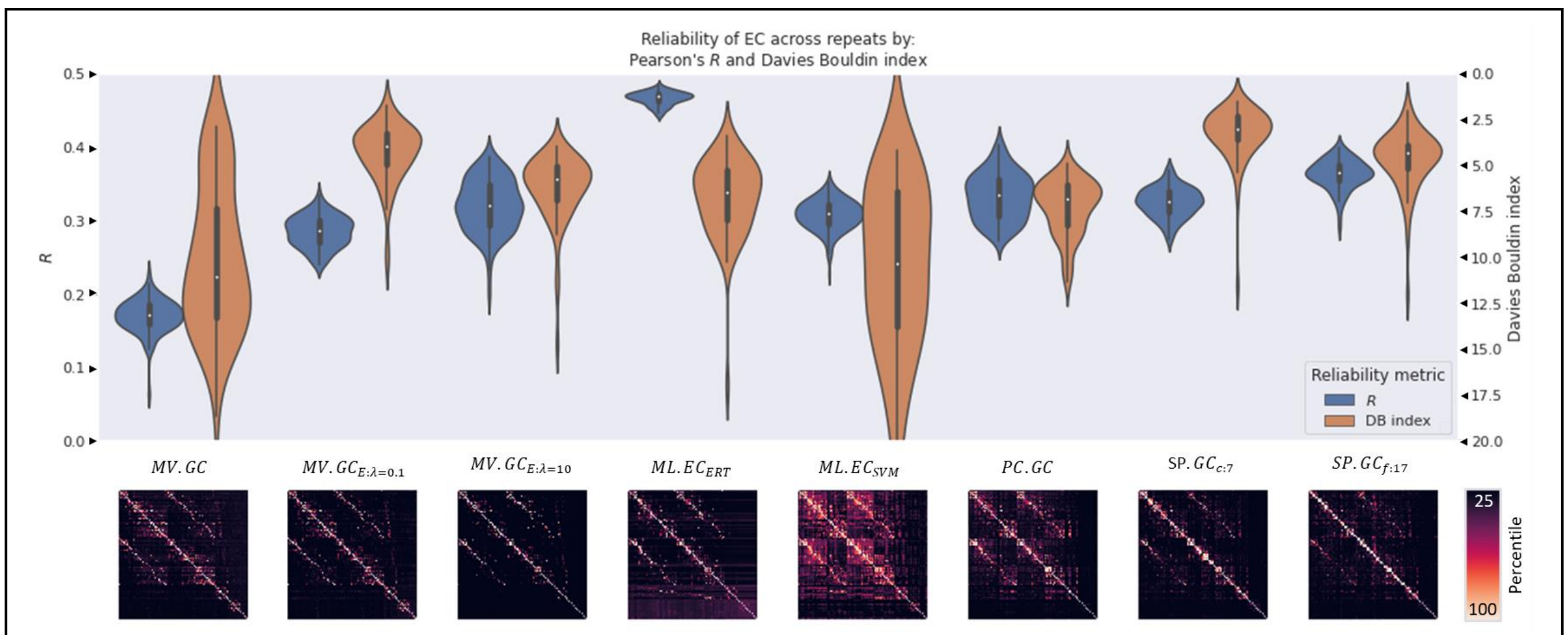

**Figure 3: Reproducibility of effective connectivity measures across repeats as measured by Pearson's *r* and Davies Bouldin Index**. The distribution of reproducibility as measured by Pearson's *r* and DB clustering score for every pair of EC matrices in the four-scan set per subject is shown. Superior reproducibility for both scores is found at the top of the graph's Y axes. The mean EC matrix for each measure is displayed under the x axis.

Full results of the GC comparison are shown in **Supplementary Figure S2.** The stability of the $SP.GC_{f:17}$ projection using short timeseries has potential for tangible benefit to analyze data from studies with limited fMRI acquisition duration.

The distributions of Pearson's *r* and the clustering score for each of the EC metrics are shown in **Figure 3.** This view manifests the clearly superior performance of $ML.EC_{ERT}$ connectivity along the Pearson's *r* metric as well as the reasonable clustering performance. Below each measure is the mean connectivity matrix across all subject scans for each measure. The ordering of the anatomy and RSNs across the columns and rows within each matrix is detailed in **Supplementary Figure S1**.

### 3.2 Experiment 2: Comparison of predictability of individuals' traits

#### 3.2.1 Comparison of trait predictability using the proposed functional connectivity measures

The ability of the connectivity metrics to predict subject level traits was evaluated in the second experiment. The most reproducible FC proposed measures ($ML.FC_{ERT}$ and $ML.FC_{XGB}$) were compared to traditional methods of FC: correlation and partial correlation-based connectivity (**Figure 4**). We first performed the univariate analysis of effect size per FC edge as described in **Section 2.4.1**. **Figure 4A** shows the effect size of the top 50 edges in predicting mean arterial pressure, **Figure 4B** shows the effect size of the top 50 edges in predicting stress, and **Figure 4C** shows the effect size of the top 50 edges in predicting fluid intelligence. The results indicate that the effect sizes of the different methods are comparable for the different predictions except for fluid intelligence, where Correlation and $ML.FC_{ERT}$ gave superior performance.

Next, a multivariate analysis was performed (**Section 2.4.2)** in which multiple edges were combined into one model to quantify whether the edges contain complementary information and determine which measures contain the largest total information about the prediction targets. **Figure 4D** shows the $R^2$ on the held-out test data. For all three targets, multivariate connectivity from $ML.FC_{XGB}$ gave the top performance, followed closely behind by Partial correlation. $ML.FC_{ERT}$ and correlation were distant 3rd

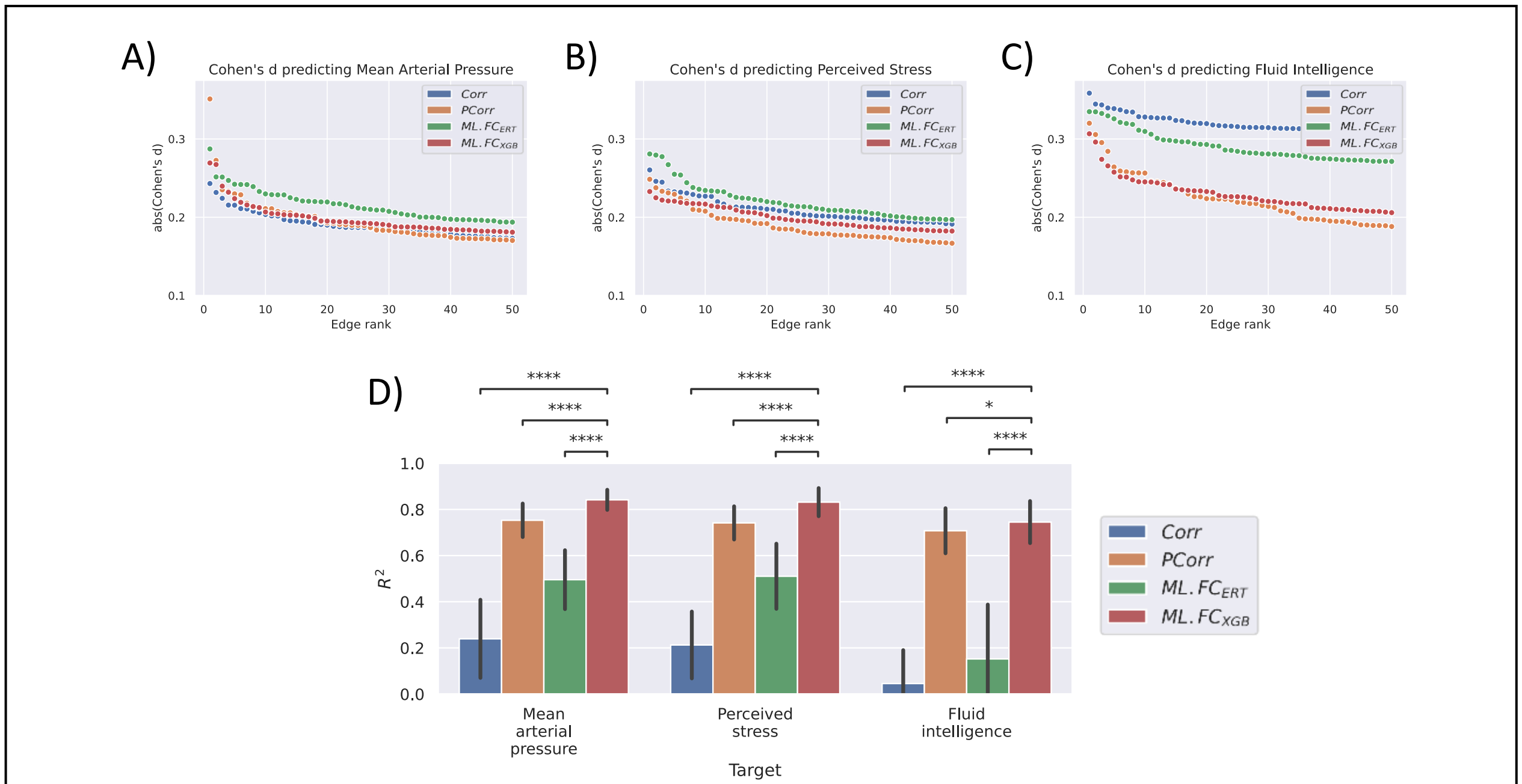

**Figure 4: Predictive ability of FC measures.** (A) shows the effect sizes of the most reproducible FC measures when used to regress a physiologic target: mean arterial blood pressure. (B) shows the effect sizes of the most reproducible FC measures when used to regress a physiologic and cognitive target: stress. (C) shows the effect sizes of the most reproducible FC measures when used to regress a cognitive target: fluid intelligence. (D) Shows the performance (as measured by $R^2$) of a model trained using the top edges ± the 95% CI for each FC connectivity method over 10 outer random permutations of the data and 10 inner cross validation folds. The top edges used were all edges from the univariate analysis of the training set with p≤0.05. Significant differences from top performer were calculated with a Bonferroni corrected one-sided t test. p>0.05, p<0.05, p<0.01, p<1e-3, and p<1e-4 are indicated with ns, *, **, ***, or **** respectively.

and 4th place finishers.

Using the elastic net predictions themselves as input to an LME model allowed us to determine which predictions were complementary (**Table 3**). Our definition of complementarity here stems from the concepts outlined in **Section 2.4.3**, where the ability of the model to predict from combinations of EC and FC data is analyzed. Complementarity then, is defined as having independent predictive power that is not found in other methods though the magnitude of appropriately regularized coefficients. In this case we observed that combining the complimentary features did not provide a statistically significant improvement in prediction accuracy. $ML.FC_{XGB}$ contained the information present in the other measures in all cases except mean arterial pressure. For that target, $ML.FC_{XGB}$ and partial correlation contained complimentary information. The results from **Figure 4** and **Table 3** suggest that the boosting method of $ML.FC_{XGB}$ is particularly well-suited to regularize and discover a stable set of connectivity features, perhaps because of its multiple-bagging approach used to handle multicollinearity.

### 3.2.2 Comparison of trait predictability using proposed effective connectivity measures

The predictive ability of the different *effective* connectivity metrics was evaluated by comparing traditional EC measures of GC, regularized GC with an elastic penalty, and $PC.GC$ to the most reproducible proposed metrics, $ML.EC_{ERT}$ with the extremely random trees predictor and $SP.GC$. **Figure 5A-C** shows the results from the univariate analysis (**Section 2.4.1**). We observed that the highest Cohen's d was attained for connections measured with $PC.GC$ (red) and $ML.FC_{ERT}$ (green) followed by $SP.GC$ (purple). **Figure 5D** shows the performance of multivariate predictive models trained on a set of all edges with univariate p-value≤0.05 (**Section 2.4.2**). The proposed $SP.GC$ method (purple) and $PC.GC$ (red) explained the most variance, followed by $ML.EC_{ERT}$ (green).

**Table 3: Complementarity of EC and FC measures.** For each set of EC and FC measures, the predicted scores from the elastic-net predictions were fed into a secondary LME model with subject-specific intercepts and group level slopes. This tested for complimentary information contained in the predictions. Those predictions using connectivity metrics were considered to contain complimentary information if the fitted coefficient had a p-value≤0.05 and a magnitude greater than or equal to 10% of the maximum coefficient magnitude. The coefficients are displayed below, with p values in parentheses. Complimentary sets of connectivity features are indicated with boldface.

| | Functional connectivity measure | | | |
|---|---|---|---|---|
| Prediction Target | $Corr$ | $PCorr$ | $ML.FC_{ERT}$ | $ML.FC_{XGB}$ |
| Mean arterial pressure | −0.02 (0.00) | **0.10** (**0.00**) | 0.02 (0.07) | **0.92** (**0.00**) |
| Perceived stress | −0.02 (0.05) | 0.06 (0.00) | 0.01 (0.14) | **0.97** (**0.00**) |
| Fluid intelligence | −0.02 (0.00) | 0.00 (0.94) | −0.01 (0.01) | **1.03** (**0.00**) |

| | Effective connectivity measure | | | |
|---|---|---|---|---|
| Prediction Target | $MV.GC$ | $MV.GC_{E:\lambda=0.1}$ | $PC.GC$ | $SP.GC_{f:17}$ |
| Mean arterial pressure | −0.05 (0.17) | **0.74** (**0.00**) | **0.08** (**0.00**) | **0.24** (**0.00**) |
| Fluid intelligence | −0.35 (0.49) | **0.31** (**0.00**) | **0.52** (**0.00**) | **0.24** (**0.00**) |
| Perceived stress | **0.24** (**0.00**) | **0.22** (**0.00**) | **0.63** (**0.00**) | **0.28** (**0.00**) |

The multivariate models (**Figure 5D**) revealed that the $SP.GC$ set of edges (purple) tended to contain more total information than the connections computed with the remaining connectivity methods and achieved the highest performance predicting arterial pressure and stress. In close second place was the $PC.GC$ connectivity measure (red), which achieved the highest performance predictive fluid intelligence.

A multi-input analysis (**Section 2.4.3)** was performed to test complementarity across the EC measures using an LME model. From the results in **Table 3** (bottom), we observe that $SP.GC$, $PC.GC$, and $MV.GC_{E:\lambda=0.1}$ contained complimentary information for all prediction targets. However, combining complimentary features did not provide a statistically significant improvement for target prediction. This suggests that the regularizing causal measures, via a low dimensional projection or an elastic penalty, extract different information. Furthermore, the cross-prediction comparison indicates that the single connectivity feature most apt to make a given prediction is somewhat problem specific, but $SP.GC$ and $PC.GC$ are well-suited to the variety of prediction tasks examined here.

## 4 Discussion

Among *functional connectivity* measures, $ML.FC_{XGB}$ had the highest reproducibility across most metrics. The overall predictive power to predict mean arterial pressure, stress, and fluid intelligence using the multivariate models was also highest for $ML.FC_{XGB}$. Additionally, our separate analysis of complementarity using second level LME models revealed that the $ML.FC_{XGB}$ approach contained most of the information present in the other FC measures. These results suggest that $ML.FC_{XGB}$ be used as the functional connectivity metric of choice on larger datasets. If there is not enough data to effectively fit the $ML.FC_{XGB}$ across subjects, then we recommend the use of partial correlation based functional connectivity, which also performed well and just behind $ML.FC_{XGB}$.

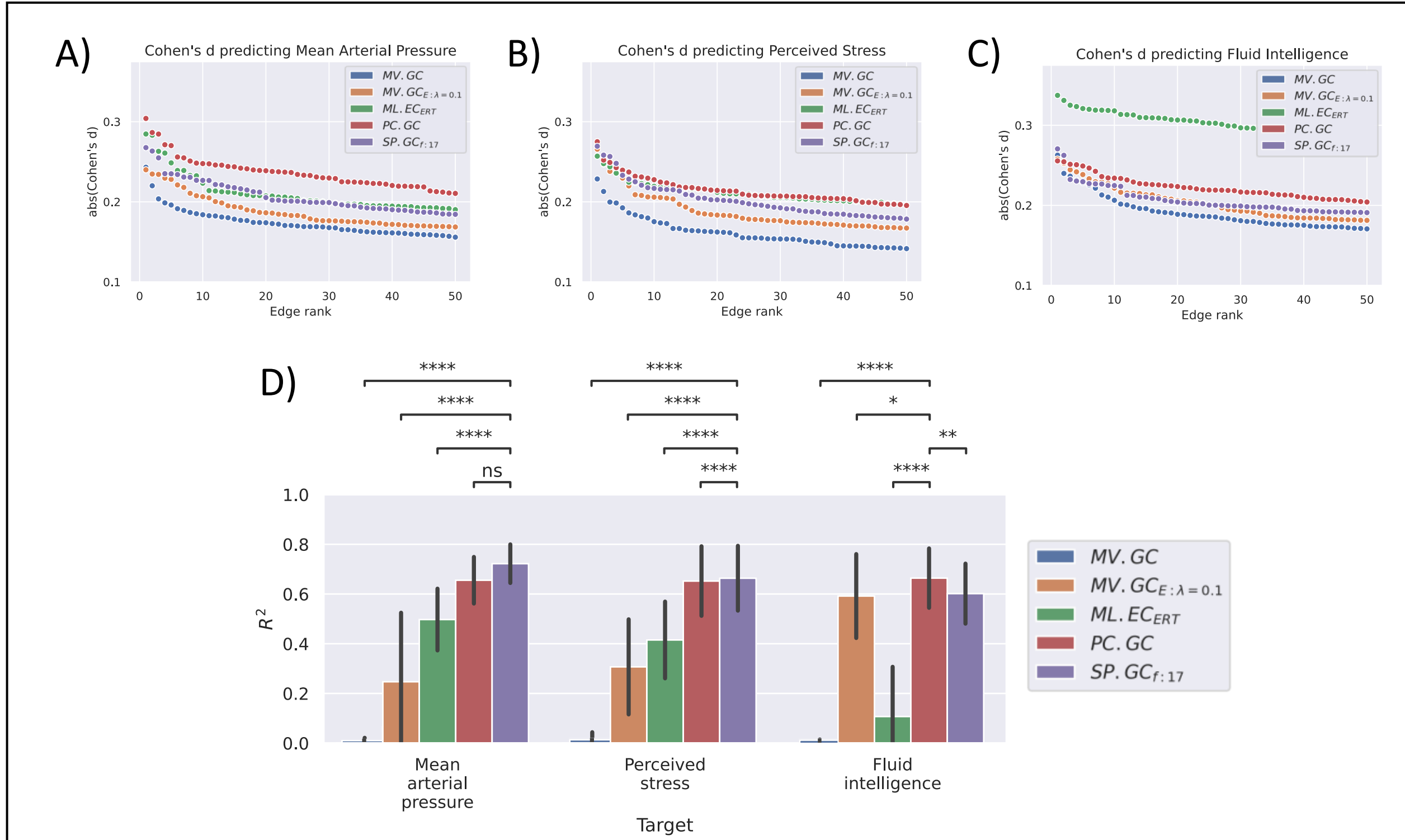

**Figure 5: Predictive ability of EC measures**. (A) shows the effect sizes of the most reproducible EC measures when used to predict a physiologic target: mean arterial blood pressure. (B) shows the effect sizes of the most reproducible EC measures when used to predict a physiologic and cognitive target: stress. (C) shows the effect sizes of the most reproducible EC measures when used to predict a cognitive target: fluid intelligence. (D) Shows the performance (as measured by $R^2$) of a model trained using the top edges ± the 95% CI for each EC connectivity method over 10 outer random shufflings of the data and 10 inner cross validation folds. The top edges used were all edges from the univariate analysis of the training set with p≤0.05. Significant differences from top performer were calculated with a Bonferroni corrected one-sided t test. p>0.05, p<0.05, p<0.01, p<1e-3, and p<1e-4 are indicated with ns, *, **, ***, or **** respectively.

Among *effective connectivity* measures, $ML.EC_{ERT}$ performed better in reproducibility than other measures, but suboptimally in the predictive power analysis, suggesting the presence of strong intra-edge correlation and redundant (collinear) information. However, $SP.GC$ performed second highest in reproducibility and was also the *most* predictive connectivity feature set in two of our three prediction models. Our analysis of reproducibility using fractional timeseries further revealed that $SP.GC_{f:17}$ (a specific formulation of $SP.GC$) achieved high reproducibility even when applied to a small portion of the fMRI timeseries (**Supplementary Figure S2**). Our complementarity analysis using LME models, showed that each of the top EC methods contained information complimentary to each other. Collectively these results indicate that: (1) the incorporation of a structural prior to the $SP.GC_{f:17}$ measure appropriately constrained the GC score with higher predictive power and reproducibility than a standard PCA projection, and (2) the $SP.GC_{f:17}$ measure may be recommended as a method for effective connectivity measurement. If appropriate dMRI priors are unavailable, we recommend either (1) $ML.EC_{ERT}$ to capture nonlinear EC connectivity with high reproducibility, or (2) $PC.GC$ if predictive power is prioritized over reproducibility.

Prior literature focused on the reproducibility of ***functional connectivity***, using Pearson's *r* or partial correlation (Andellini et al., 2015; Fiecas et al., 2013; Geerligs et al., 2017; Guo et al., 2012; Liao et al., 2013; Noble et al., 2019; Noble et al., 2017b; Noble et al., 2017a; Pannunzi et al., 2017; Termenon et al., 2016; Wang et al., 2017). Measures of reproducibility can be highly confounded (usually inflated) by motion, therefore the selection of subjects was aimed to minimize this confound (Noble et al., 2019). Our estimates of reproducibility of FC measures is somewhat lower than prior reports in the literature (Noble et al., 2017a) which is likely a result of our strict motion thresholds limiting the inflation of regional correlation from motion. This is supported by previous reviews of connectivity (Noble et al., 2019; Noble et al., 2017b). The aforementioned prior research characterizes the reproducibility of correlation and partial correlation well, but used a limited set of reliability metrics, typically intraclass correlation coefficient (ICC) or $R^2$ (Noble et al., 2017a; Termenon et al., 2016; Waller et al., 2017). This study complements the prior work in three ways. First, we provide a characterize *reproducibility* using a multitude of metrics (linear, nonlinear, and clustering). Second, we characterize *predictive power* using several metrics (ICC, $R^2$, cosine similarity, DB score, and accuracy across 3 relevant neurophysiologic targets). Finally, we *propose new measures* of connectivity, quantify their reproducibility, and compare them to traditional measures. These proposed measures proved to be the most reproducible and contained the greatest predictive power of the FC measures analyzed. The measure proposed by Murugesan et al (Murugesan et al., 2020) is a special case of the generalized framework for $ML.FC$ proposed in this work. Our framework generalizes across model types (SVM, ERT, and XGB), hyperparameter optimization, develops machine learning connectivity measures for both FC and EC, and adds regularization priors, while the measure in (Murugesan et al., 2020) only pertains to FC. Additionally, we evaluate the reproducibility of the proposed models, which is absent from such prior literature.

Examinations of the reproducibility of ***effective connectivity*** have been limited to bivariate GC (Fiecas et al., 2013), which is a Granger causal estimate using only pairs of regional timeseries, rather than the more comprehensive multivariate estimates using all regional timeseries employed in this study. This research drives the study of the reproducibility of EC measures beyond bivariate GC analysis. Furthermore, although many studies do not characterize performance, reproducibility is necessary but insufficient quality of a desirable neuroimaging predictor (Noble et al., 2017b; Noble et al., 2017a; Termenon et al., 2016; Waller et al., 2017). Our analysis of predictive power over multiple tasks addresses this gap as well.

When comparing EC and FC measures, the best FC measures had slightly higher reproducibility and predictive power than the best EC measures. However, FC and EC capture different connectivity information and should be considered complimentary rather than competing. FC, for example, may capture longer range interactions, while shorter direct interactions may be better captured by EC. If interpretability is the most important criterion, then we recommend the use of $SP.GC_{f:17}$. If predictive performance is the most important criterion, then we recommend $ML.FC_{XGB}$. If both are equally important, we recommend combining the two, using the top *X* percent of informative edges (from a univariate analysis) from each measure, where *X* is determined based on the number of samples in the study so as to make the ratio of features to samples tractable to train a predictive model without overfitting.

There are several limitations of this research. *First*, comparison of reproducibility across different connectivity types can be problematic when the distribution of recovered connectivity values is concentrated to a few values. For example, a calculated GC connectivity with an elastic penalty where the $\lambda$ penalty is extremely high can have 99% values of 0 connectivity, and the remaining nonzero connections can be uniformly distributed on a logarithmic scale within the range $[0, \infty]$. On the other hand, correlative connectivity values are within the range [−1,1] and tend to be more normally distributed. By using multiple metrics of comparison including several metrics of both reproducibility and predictive power, we increase confidence in the relative usefulness of the analyzed connectivity measures. However, additional reproducibility metrics could be explored. For example, summary graph connectivity measures may exhibit useful predictive properties and reliability, but the differing sparsity between methods would need to be addressed to make such a comparison. The *second* limitation is that metrics of connectivity and reproducibility can be dependent upon preprocessing decisions, number of subjects, and number of replicates per subject. Future studies could explore the dependency of FC and EC measures to denoising, global signal regression, and choice of atlas, as well as look at similarity among the new and classic methods in terms of edge similarity, measures of graph similarity, and how these change with more noise added. When additional large-scale datasets with multiple replicates per subject become available, further work on reproducibility across more replicates and subjects would also be valuable and welcome. Finally, the *third* limitation is that, in an effort to focus on the ability of measures to quantify biological variability and not merely quantify motion artifacts, we purposefully chose a sample of subjects with exceptionally low motion. As researchers implement better practices which yield lower and lower intrinsic motion and as motion suppression methods improve, our choice becomes more and more reasonable. However we emphasize that our results pertain to studies with low motion, long scan times, and high temporal resolution. Furthermore, we chose to use all available data per subject, including 4 longer-duration scans per subject to make each prediction, which can increase the estimated $R^2$. The real-world ability to predict in these domains we would expect to be roughly uniformly lower for all methods, however, this study quantifed the *relative* ability of the models to make predictions *within* a dataset with lower motion. A full study estimating the real-world power of these methods is beyond the scope of this work, which proposes and tests the internal validity of novel algorithms. However, we have also successfully made realistic predictions with these mesures in a Parkinson's disease predictive modeling system, and direct the reader to that body of work (Mellema et al., 2023).

## 5 Conclusions

This study proposes a functional connectivity metric (*ML.FC*) and an effective connectivity metric (*ML.EC*) that efficiently capture nonlinear associations between brain regions. This study also proposes a connectivity metric (*SP.GC*) that encourages the connectivity recovered from fMRI to respect underlying biological structural connectivity and efficiently measures causal associations across all brain regions, which provides researchers new capabilities for connectivity analysis. This study compared the proposed measures to traditional ones using quantitative *reproducibility* metrics and by quantifying their capacity to make accurate *predictions* of traits of individual subjects to show the internal validity of these proposed metrics. This included a physiologic trait, cognitive trait, and combined physiologic and cognitive trait. The proposed measures produced higher measures of reproducibility and were found to be more predictive across the traits. Based on the study results, two of the proposed methods: $ML.FC_{XGB}$ and $SP.GC_{f:17}$ are recommended as the connectivity measures of choice for functional and effective connectivity, respectively. The contributions of this work hold potential to further the development of tools to characterize the human connectome in health and disease and make meaningful individualized predictions of neuropsychological and neurobiological states.

## 6 Acknowledgements

Special thanks Dr. Prapti Modi, PhD, and Dr. Daniel Heitjan, PhD, for providing additional feedback and editing for this manuscript. Cooper Mellema was supported by NIH NINDS F31 fellowship NS115348. Albert Montillo was supported by NIH NIA R01AG059288, NIH NCI U01 CA207091, NIH NIGMS R01GM144486, the King Foundation, and the Lyda Hill Foundation. The authors declare no competing interests.

## 7 Materials and ethics statement

To facilitate reuse and extension, the authors are pleased to provide full source code for the connectivity measures and manuscript analyses at: (https://git.biohpc.swmed.edu/s169682/ CausalMeasures). Datasets used for this study are publicly available at (http://www. humanconnectomeproject.org/). The data used for this study data was gathered with written informed consent from all participants. Furthermore, this manuscript follows NIH policy on the inclusion of women and members of racial and ethnic minority groups in funded clinical research. Participant data was collected and handled in accordance with NIH guidelines, HIPPA guidelines, local statutory requirements, and handled with the principles embodied in the Declaration of Helinski.

## 9 Supplement

### 9.1.1 Abbreviations

A summary table of abbreviations used throughout this manuscript is described in **Supplementary Table S1.** Each connectivity measure is designated as {Measure technique}.{Measure category} with modeling specifics as a subscript.

**Table S1**: Table of abbreviations used throughout the manuscript.

| Abbreviation | Description |
|---|---|
| **fMRI** | Functional Magnetic Resonance Imaging |
| **BOLD** | Blood-oxygen Level dependent |
| **dMRI** | Diffusion Magnetic Resonance Imaging |
| **FC** | Functional Connectivity |
| **EC** | Effective Connectivity |
| **GC** | Granger Causality |
| **MVAR** | Multivariate Autoregressive |
| **HCP** | Human Connectome Project |
| **BOHB** | Bayesian Optimization Hyper Band |
| **TR** | Repetition Time |
| **AIC** | Akaike Information Criterion |
| **PCA** | Principal Component Analysis |
| **SC** | Structural Connectivity |
| **ROI** | Region Of Interest |
| **RSN** | Resting State Network |
| **AAL** | Automated Anatomical Labeling |
| **APA** | Asian or Pacific-Islander American |
| **Corr** | Correlation functional connectivity |
| **PCorr** | Partial Correlation functional Connectivity |
| $SG.FC$ | Spectral Granger causality Functional Connectivity measure |
| $ML.FC_{M}$ | Machine-Learning Functional Connectivity measure using machine learning model M |
| $ML.EC_{M}$ | Machine-Learning Effective Connectivity measure using machine learning model M |
| $ERT$ | Extremely random trees regressor |
| $SVM$ | Support vector machine regressor with a radial basis function kernel |
| $XGB$ | Extreme gradient boosting regressor |
| $ML.FC_{ERT}$ | Machine Learning Functional Connectivity measure with an ERT regressor |
| $ML.FC_{SVM}$ | Machine Learning Functional Connectivity measure with an SVM regressor |
| $ML.FC_{XGB}$ | Machine Learning Functional Connectivity measure with an XGB regressor |
| $ML.EC_{ERT}$ | Machine Learning Functional Connectivity measure with an ERT regressor |
| $ML.EC_{SVM}$ | Machine Learning Functional Connectivity measure with an SVM regressor |
| $MV.GC$ | Multivariate autoregressive Granger Causality measure with a linear model |
| $MV.GC_{E:\lambda=0.1}$ | Multivariate autoregressive Granger Causality measure with an elastic model with regularization term $\lambda = L_1 = L_2$ penalty of 0.1 |
| $MV.GC_{E:\lambda=10}$ | Multivariate autoregressive Granger Causality measure with an elastic model with regularization term $\lambda = L_1 = L_2$ penalty of 10 |
| $SP.GC$ | Structurally Projected Granger Causality measure |
| $SP.GC_{c:7}$ | Structurally Projected Granger Causality measure with a coarse SC prior with 7 bilateral ROIs |
| $SP.GC_{f:17}$ | Structurally Projected Granger Causality measure with a fine SC prior with 17 bilateral ROIs |
| $PC.GC$ | Principal Component analysis (PCA) projected Granger Causality measure |
| **ICC** | Two-way random single score Intraclass Correlation Coefficient |
| **DB** | Davies-Bouldin Index |
| **RMSE** | Root Mean Square Error |
| **FDR** | False Discovery Rate |

### 9.1.2 Functional connectivity definitions

The FC measures used include Pearson's r, partial correlation, and spectral Granger causality. These measure a degree of linear association between two mean regional timeseries. Pearson's r measures the direct linear correlation between two sets of data – normalizing covariance between timeseries $X_1$ and $X_2$ by their standard deviations $\sigma_{X_1}$ and $\sigma_{X_2}$. The equation for Pearson's R is:

$$r = \frac{cov(X_1, X_2)}{\sigma_{X_1}\sigma_{X_2}}. \quad \textbf{(S1)}$$

The calculation of partial correlation is closely related to Pearson's $r$. Partial correlation measures the degree of linear association between mean regional timeseries with confounding variables removed. The partial correlation between timeseries $X_1$ and $X_2$ while normalizing for $n$ controlling variables $X_{n\notin 1,2}$ (in the case of neuroimaging, all other regional timeseries) is defined as the correlation between the residuals $\epsilon_{X_1}$ and $\epsilon_{X_2}$ for linear models predicting $X_1$ or $X_2$ from $X_{n\notin\{1,2\}}$. This formulation is presented in **Supplemental Equation S2**, where $f$ is a linear model predicting $X_i$ from $X_{n\notin\{1,2\}}$.

$$\rho = \frac{cov(\epsilon_1, \epsilon_2)}{\sigma_{\epsilon_1}\sigma_{\epsilon_2}}, \qquad \epsilon_i = X_i - f\left(X_{n\notin\{1,2\}}\right) \quad \textbf{(S2)}$$

Timeseries can additionally be represented in frequency space, and other means of calculating FC can involve comparisons in this domain. Particularly, we employ a mean across spectral densities, also known as spectral Granger Causality, to evaluate functional connectivity in the frequency domain. By utilizing the Geweke formulation of Granger Causality in the frequency domain, we can express a measure of spectral density transfer between timeseries at a specific frequency (Mingzhou Ding, Yongheng Chen, Stephen L. Bressler, 2006). The transfer is then symmetrized and averaged over physiologic frequencies of 0.02-0.15 Hz for greater solution stability. For further details of spectral Granger Causality see Ding et al. (Mingzhou Ding, Yongheng Chen, Stephen L. Bressler, 2006).

### 9.1.3 Feature importance calculation

For the machine learning based FC measures described in **Section 2.1.1**., details of the feature importance calculation and hyperparameters selection are provided here. *Feature importance* is calculated as follows for each model. For the ERT model, we use the Gini importance as this feature importance weight. For the SVM, we use the covariate weight. And for the XGB model, we use the Gini importance weighted by the number of samples routed through any given decision node. *Hyperparameters* are determined as follows. A two-step approach is used to fit the XGB models. In the first step, a group-level model to predict each brain region's activity from data pooled from all subjects is fit and hyperparameter optimized with a Bayesian Optimization Hyper Band (BOHB) search. This identifies reasonable values for a few general hyperparameters: the maximum depth, learning rate, L1 and L2 regularization, and minimum loss per split. The hyperparameter range searched over is outlined in **Table S2** below. For the second step, we fit a subject-specific model for EC calculation using the hyperparameters found in step 1. This is done by training the model with an equal weight on the target subject, as well as a regularization prior consisting of all other subjects (e.g., the model must fit both the specific subject as well as the population of subjects, with an equal weighting on each). Tree-based models are not readily adapted for incremental learning, so rather than tuning a group-level model on subject specific data, we use a combined dataset as recommended by the XGB developers (Chen and Guestrin, 2016). The ERT and SVM models did not benefit from a hyperparameter search relative to their default parameters (as tested on HCP data NOT used in training, validation, or testing), and the default parameters were used.

**Table S2: Hyperparameter optimization of XGBoost.** This figure shows the hyperparameter ranges searched over in our XGBoost hyperparameter optimization. The hyperparameter name is in the left column, distribution it was sampled from in the center column, and range of the hyperparameter in the right column.

| *Hyperparameter* | *Distribution* | *Range* |
|---|---|---|
| Depth | Uniform | [3,9] |
| Learning rate | Loguniform | [0.1,100] |
| Booster | Choice | [dart, tree] |
| Tree | Choice | [aprx, hist, auto, exact] |
| Gamma | Loguniform | [0.01, 10] |
| Eta | Uniform | [0,1] |
| Min weight | Uniform | [0,2] |
| Max delta | Uniform | [0,2] |
| Subsample | Uniform | [0.1,1] |
| Alpha | Loguniform | [1e-6,1e2] |
| Lambda | Loguniform | [1e-6,1e2] |
| Tree node sampling | Uniform | [0.4,1] |
| Tree level sampling | Uniform | [0.4,1] |
| Number of estimators | Uniform | [50,500] |

### 9.1.4 Structural connectivity prior informed sparse principal components analysis equation

In **Equation 2**, $\vec{v_i}^*$ is the structural-connectivity-informed sparse principal component. $\vec{v_i}$ is the $i^{th}$ principal component as we iteratively find each sparse component. $\boldsymbol{P}$ is the prior matrix derived from the structural connectivity. $\boldsymbol{C}$ is the covariance matrix between mean regional timeseries. $\theta$ is a scalar weight which governs the strength of the prior **P**. The prior is constructed from the logarithm of an intermediate matrix $S_{i\cap j}$, which is calculated from the structural connectivity matrix. $S_{i\cap j}$ is a block-averaged version of the original structural connectivity matrix, where each row $i$ corresponds to a specific region of interest (ROI), and each column corresponds to a larger network $j$. These larger networks are known resting state networks (RSNs) comprised of multiple ROIs. Each element $i, j$ represents the intra and inter RSN connectivity. $S_{i\cap j}$ is the summed set of all streamlines that connect the RSN that ROI $i$ is a subset of to RSN $j$. For example, $S_{1\cap 1}$ corresponds to the sum of all streamlines which connect sub-regions of RSN 1, representing intra-RSN connectivity, while $S_{1\cap 2}$ corresponds to the sum of all streamlines which connect any subregion of RSN 1 to RSN 2. Given say 100 regions and 10 RSNs, matrix $\boldsymbol{P}$ would have dimensions $N_{RSN} \times N_{ROI} = 10 \times 100$, where each row describes the inter-RSN connectivity of that ROI. Each row of the SC-derived prior matrix $\boldsymbol{P}$ corresponds to a 'subnetwork' of the structural connections, and each column describes which subnetwork a specific ROI belongs to. The prior matrix must have fewer RSNs than the number of regional timeseries to reduce the dimensionality in this formulation of constrained sparse PCA, where the number of components is prespecified by the prior dimension. The regularized, sparse principal components will be used in the low-dimensional GC analysis, and the resulting GC scores per region can then be calculated in **Algorithm 3**.

### 9.1.5 Connectivity matrix organization

**Supplementary Figure S1** illustrates how the connectivity matrices of **Figure 2** and **Figure 3** are organized. **Figure S1** shows how the rows and columns are organized anatomically by hemisphere and by resting state network (RSN). The cortical regions are sorted first by left/right hemisphere and then by RSN, while the subcortical and cerebellar regions are sorted first by subcortical region and then by left/right hemisphere.

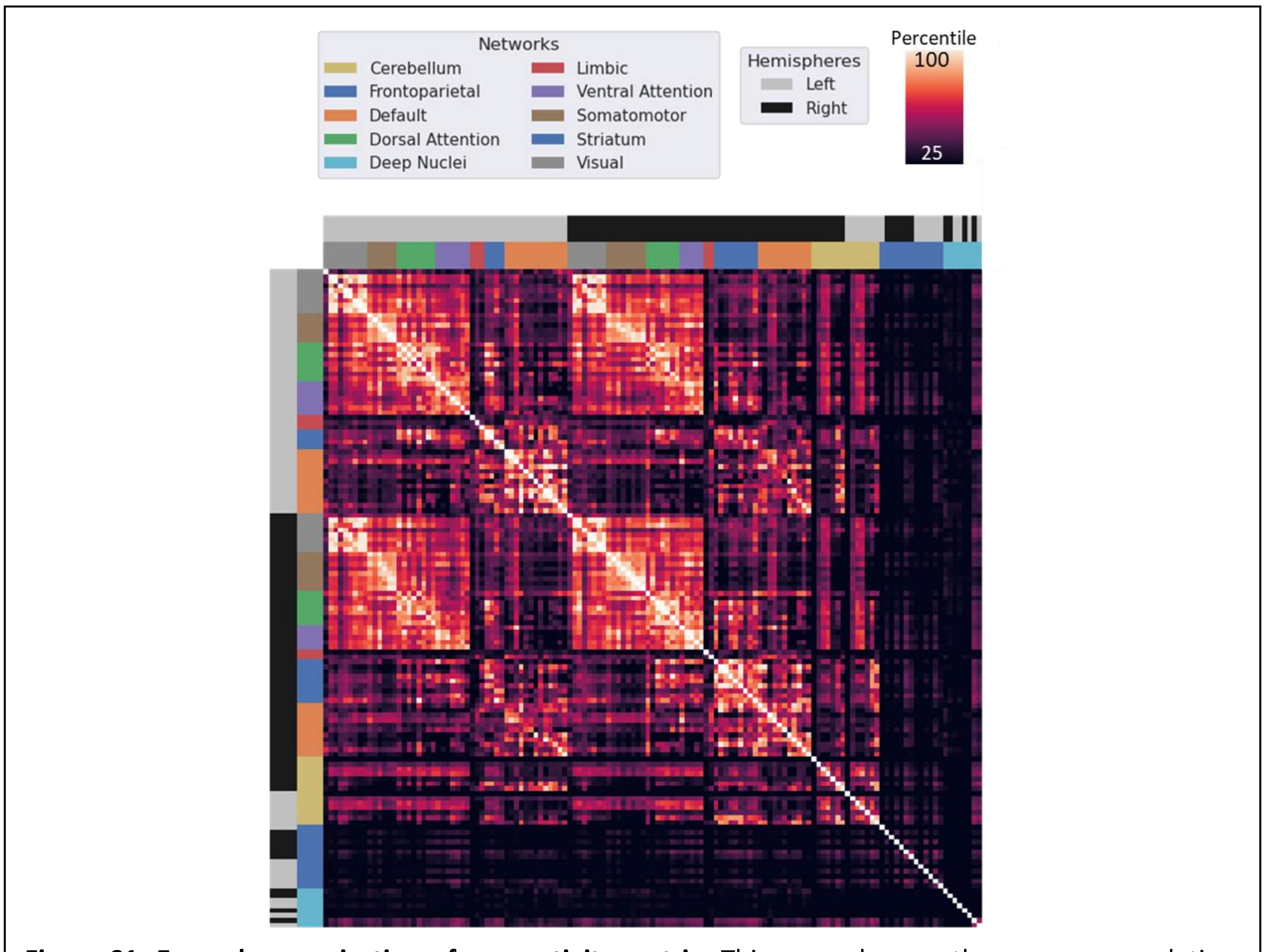

**Figure S1: Example organization of connectivity matrix**. This example uses the average correlation matrix across all 100 subjects and is organized first by hemisphere and secondarily by RSN. The values within the connectivity matrix are normalized through percentile scaling.

### 9.1.6 Effective connectivity stability

To further differentiate between the different causal connectivity metrics, we quantified the stability of these measures as a function of the portion of the timeseries considered. An ideal metric is highly reproducible even when calculated on a small fraction of the timeseries. **Supplementary Figure S2** shows the performance of the top EC measures: $\mathrm{MV.GC}_{\mathrm{E:\lambda=0.1}}$, $\mathrm{PC.GC}$, and $\mathrm{SP.GC}_{\mathrm{f:17}}$. The reliability of and $\mathrm{SP.GC}_{\mathrm{f:17}}$ is significantly higher than the other methods, and maintains this higher reproducibility across all lengths of time analyzed. Notably, $\mathrm{SP.GC}_{\mathrm{f:17}}$ using 50% of the initial timeseries has a higher reproducibility than the $\mathrm{MV.GC}_{\mathrm{E:\lambda=0.1}}$ or $\mathrm{PC.GC}$ methods using 100% of the timeseries.

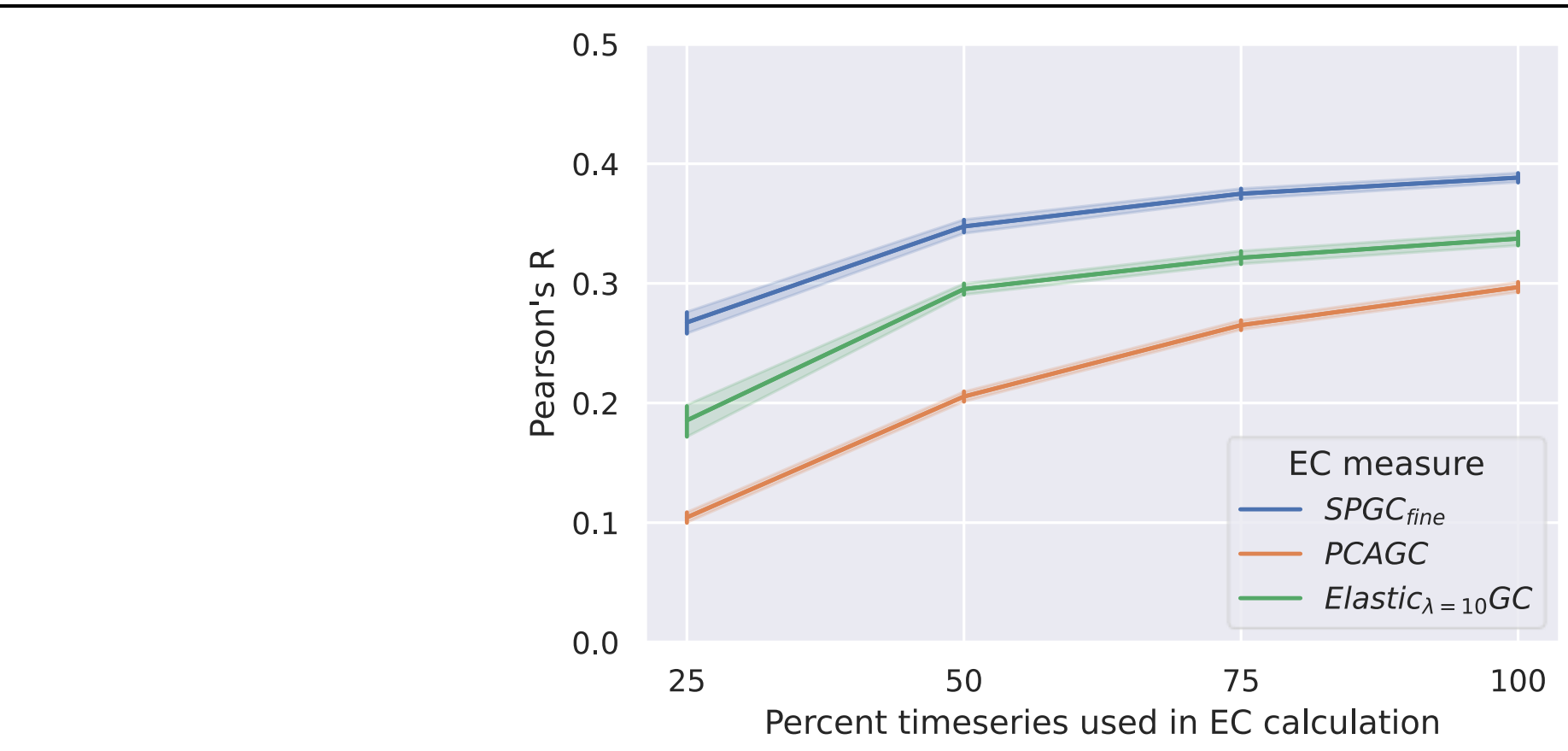

**Figure S2: Reliability of effective connectivity using a partial timeseries**. The reproducibility as measured by average Pearson's *r* between every set of EC matrices per subject is shown versus the percent of the initial timeseries used. For example, the point "25% of the timeseries used" corresponds to the Pearson's R on the calculated EC or FC matrix across each subjects' 4 repeat scans when only the first 25% of the original timeseries was used. The 95% CI is shown as error bars and shading around each curve.